\documentclass[aps,showpacs,twocolumn,twoside,groupedaddress]{revtex4}
\usepackage{graphicx}
\usepackage{epsfig}
\usepackage{epsf}
\usepackage{amssymb}
\usepackage{amsmath}

\begin{document}

\title{Single Photon Transport through an Atomic Chain Coupled to a One-dimensional Nanophotonic Waveguide}

\author{Zeyang Liao$^{1,2}$\footnote{zeyangliao@physics.tamu.edu}, Xiaodong Zeng$^{1}$, Shi-Yao Zhu$^{2}$, and M. Suhail Zubairy$^{1,2}$\footnote{zubairy@physics.tamu.edu}}

\affiliation{$^1$Institute for Quantum Science and Engineering (IQSE) and Department of Physics and Astronomy, Texas A$\&$M University, College Station, TX 77843-4242, USA \\
$^2$Beijing Computational Science Research Center, Beijing 100084, China \\
}

\begin{abstract}
We study the dynamics of a single photon pulse travels through a linear atomic chain coupled to a one-dimensional (1D) single mode photonic waveguide. We derive a time-dependent dynamical theory for this collective many-body system which allows us to study the real time evolution of the photon transport and the atomic excitations. Our analytical result is consistent with previous numerical calculations when there is only one atom. For an atomic chain, the collective interaction between the atoms mediated by the waveguide mode can significantly change the dynamics of the system. The reflectivity of a photon can be tuned by changing the ratio of coupling strength and the photon linewidth or by changing the number of atoms in the chain. The reflectivity of a single photon pulse with finite bandwidth can even approach $100\%$. The spectrum of the reflected and transmitted photon can also be significantly different from the single atom case. Many interesting physical phenomena can occur in this system such as the photonic bandgap effects, quantum entanglement generation, Fano-like interference, and superradiant effects. For engineering, this system may serve as a single photon frequency filter, single photon modulation and may find important applications in quantum information.
\end{abstract}

 \maketitle

\section{Introduction}

Photons are ideal carriers for quantum information. Hence, manipulating and routing photons can have important applications in the quantum information technology which has been extensively studied \cite{Raimond2007, Pelton2002, Rempe2007, Chang2007, Hwang2009, Bermel2006}. However, the photons rarely interact with each other. Therefore, we have to resort to the photon-atom interaction to modulate and control the photons. The photon-atom coupling in the vacuum is usually very weak. We can, however, modify this coupling strength by changing the environment of the vacuums by Purcell effect \cite{Purcell1946}. Strong photon-atom interaction can be achieved by confining the photon in reduced dimensions such as in quasi 1D photonic waveguide with transverse cross sections on the order of a wavelength square \cite{Leistikow2011, Noda2007}. There are a number of systems that can act as a quasi 1D waveguide such as optical nanofibers \cite{Dayan2008}, photonic crystal with line defects \cite{Englund2007}, surface plasmon nanowire \cite{Akimov2007}, and superconducting microwave transmission lines \cite{Wallraff2004, Lalumiere2013}. This kind of waveguide is extremely interesting because it can not only enhance the interaction but also guide the photon which is important for the information transport. 

A single photon scattered by a single atom embedded in a one-dimensional waveguide has been studied in the pioneering work by Shen and Fan \cite{Shen2005a, Shen2005b}. They employed the real space description of the Dicke Hamiltonian and Bethe-ansatz to derive the stationary transport properties of a single photon-atom interaction. It shows that the photon transmission spectra can be strongly modified. In particular, a photon with frequency resonant to the two-level atom can be even completely reflected. Since then, this method has been generalized to the case for multi-level atom and multi-photon interactions \cite{Shen2007, Roy2011, Tsoi2009, Sun2014, Fang2014, Fang2015}. In the stationary calculations, the photon is assumed to be a plane wave with single frequency. However, in the real case the photon is always a pulse with a finite bandwdith. An extension of the stationary theory was developed by Rephaeli {\it et al.} to derive the perfect inverting pulse where they project the initial quantum state to the stationary eigenstates of the system and then apply a time evolution operator \cite{Rephaeli2010}. In 2011, Chen {\it et al.} illustrated a time-dependent theory for a single photon absorption by single emitter coupled to one-dimensional photonic waveguide \cite{Chen2011}. This theory can be used to study the dynamics of the photon wavepacket and the atomic excitation in real time. The spontaneous emission from a pair of two-level atoms near a nanofiber has also been studied where a substantial radiative exchange is predicted between distant atoms \cite{Kien2005}. 

The collective spontaneous emission by an ensemble of identical atoms has been extensively studied since the pioneering work of Dicke in 1954 \cite{Dicke, MacGillivray, Gross, Benedict, Skribanowitz, Scully2009, Svidzinsky2008, Svidzinsky2010}. The Dicke super-radiance provides a valuable example for studying the many-body physics of photons and atoms \cite{Rohlsberger}. The interaction between the atoms mediated by the common vacuum field can dramatically change the decay rate and the eigenspectrum of the atomic ensemble \cite{Liao2014}. The decay rate of the system can be superradiant or subradiant. The Dicke symmetric state of maximum cooperation has decay rate N times the single atom decay rate \cite{Dicke}. The emission spectrum can also be significantly altered. The stationary calculation of a single photon scattered by multiple emitters coupled to one-dimensional photonic waveguide has also been studied where it is shown how the transmission spectrum can be strongly modified by the quantum interference effects \cite{Tsoi2008, Roy2013}. However, in this analysis the collective dipole-dipole interactions between the atoms are not included. 

A general master equation for describing the interaction between a quantum system and a continuous-mode Fock state has been studied \cite{Baragiola2012}. In this paper, we use a different approach to derive a time-dependent theory of a single photon scattered by an atomic chain coupled to a 1D waveguide where the collective many-body interaction can play a crucial role. We can calculate how the atomic excitations evolves in real time and how the single photon pulse is reflected and transmitted. We also show how the reflectivity and the spectrum are modified by this collective many-body system. Many interesting physical phenomena can occur in this system such as the photonic bandgap effects, quantum entanglement generation, Fano-like interference, and superradiant effects. This system may serve as a single photon frequency filter, single photon modulation and may find important applications in quantum information.

This paper is organized as follows. In section II, we derive the time-dependent theory for a single photon transport through an atomic chain coupled to a one-dimensional photonic waveguide. In section III, we discuss the one-atom case where we derive an analytical solution which has not yet been obtained before and our result is consistent with the previous numerical results. In section III, we numerically study how a single photon can be scattered by a two-atom system and show some new features compared to one-atom case. In section IV, we study the N-atom case and show how the atomic separation and atom numbers affect the reflection and transmission of a single photon. In section V, we study the effects of spontaneous decay to the free space. Finally, we summarize the results.

\section{Theory}

\begin{figure}
\includegraphics[width=0.95\columnwidth]{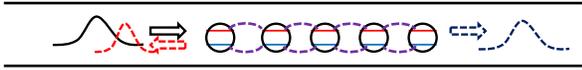}
\caption{(Color online) Single photon transport through a linear atomic chain coupled to a one-dimensional photonic waveguide. The atoms can be coupled by the waveguide vacuum modes. } 
\end{figure}

We consider a linear atomic chain coupled to a one-dimensional single-mode waveguide (Fig. 1).  We assume that the atoms are equally spaced with separation $a$ and the number of atoms is $N_{a}$. A linearly polarized single photon pulse with spectral width $\Delta\omega$ propagates through and interacts with the atomic chain. The Hamiltonian of the system in the rotating wave approximation is given by \cite{Dicke, Scully2001}
\begin{eqnarray}
H&=&\hbar(\omega_{a}-i\frac{\gamma}{2})\sum_{j=1}^{N_{a}}S_{j}^{+}S_{j}^{-}+\hbar\sum_{k}\omega_{k}a_{k}^{\dagger }a_{k} \nonumber \\ & &+\hbar\sum_{j=1}^{N_{a}}\sum_{k}(g_{k}e^{ikr_{j}}a_{k}S_{j}^{+}+g_{k}^{*}e^{-ikr_{j}}a_{k}^{\dagger }S_{j}^{-})
\end{eqnarray}
where the first term is the atomic energy including the spontaneous decay to the free space, the second term is the energy of the guided photon, the third term is the coupling between the atoms and the guided photon modes. Here $\omega_{a}$ is the transition frequency of the two-level atoms, $\gamma$ is the rate of spontaneous decay to the non-guided modes, $S_{j}^{+}=|e\rangle_j\langle g| (S_{j}^{-}=|g\rangle_j\langle e|)$ is the raising (lowering) operator of the $j$th atom, $a_{k}^{\dagger }(a_{k}^{-})$ is the creation (annihilation) operator of a photon with wavevector $k$, $g_{k}$ is the coupling strength between the atoms and the guided photon mode. $\omega_{k}$ is the angular frequency of the photon with wavevector $k$ and $r_{j}$ is the position of the jth atom. Assuming that $\omega_{a}$ is far away from the cutoff frequency of the photonic waveguide and guided photon has a narrow bandwidth, the dispersion relation for the guided photon can be approximately linearized as $\omega_{k}=\omega_{a}+(|k|-k_{a})v_{g}$ where $k_{a}$ is the wavevector at frequency $\omega_{a}$ and $v_{g}$ is the group velocity \cite{Shen2009}. 

For the single photon excitation, the quantum state of the system at any time can be expressed as 
\begin{eqnarray}
|\Psi(t)\rangle &=& \sum_{j=1}^{N_{a}}\alpha_{j}(t)e^{-i\omega_{a}t}|e_{j},0\rangle+\sum_{k}\beta_{k}(t)e^{-i\omega_{k}t}|g,1_{k}\rangle \nonumber \\ 
\end{eqnarray}
where $|e_{j},0\rangle$ is the state in which all the atoms except the $j$th atom are in the ground state with zero photon in the waveguide and the free space, $|g,1_{k}\rangle$ is the state in which all the atoms are in the ground state and one photon with wavevector $k$ is in the waveguide.

The dynamics of the system is given by 
\begin{eqnarray}
\dot{\alpha}_{j}(t)&=&-i\sum_{k}g_{k}e^{ikr_{j}}\beta_{k}(t)e^{-i\delta\omega_{k}t}-\frac{\gamma}{2}\alpha_{j}(t), \\
\dot{\beta}_{k}(t)&=&-i\sum_{j=1}^{N_{a}}g_{k}^{*}e^{-ikr_{j}}\alpha_{j}(t)e^{i\delta\omega_{k}t}, 
\end{eqnarray}
where $\delta\omega_{k}\equiv \omega_{k}-\omega_{a}=(|k|-k_a)v_{g}$ is the detuning between the atomic transition frequency and the frequency of the guided photon. 

Integrating Eq. (4) we obtain
\begin{equation}
\beta_{k}(t)=\beta_{k}(0)-i\sum_{j=1}^{N_{a}}g_{k}^{*}e^{-ikr_{j}}\int_{0}^{t}\alpha_{j}(t')e^{i\delta\omega_{k}t'}dt', 
\end{equation}
where $\beta_{k}(0)$ is the initial photon amplitude. 
On substituting from Eq. (5) into Eq. (3) we get the dynamic of the atomic excitations
\begin{eqnarray}
\dot{\alpha}_{j}(t)=&-&i\sum_{k}g_{k}e^{ikr_{j}}\beta_{k}(0)e^{-i\delta\omega_{k}t}-\frac{\gamma}{2}\alpha_{j} \nonumber \\
&-&\sum_{l=1}^{N_{a}}\sum_{k}|g_{k}|^{2}e^{-ik(r_{j}-r_{l})}\int_{0}^{\infty }\alpha_{l}(t')e^{i\delta\omega_{k}t'}dt'e^{-i\delta\omega_{k}t} \nonumber \\
\end{eqnarray}

For a long 1D waveguide we can replace the summation over $k$ by integration
\begin{equation}
\sum_{k}\rightarrow \frac{L}{2\pi}\int_{-\infty }^{\infty }dk, 
\end{equation}
where $L$ is the quantization length in the propagation direction. 

According to the Weisskopf-Wigner approximation, on summing over $k$ and integrating over $t'$, Eq. (6) becomes
\begin{equation}
\dot{\alpha}_{j}(t)=b_{j}(t)-\sum_{l=1}^{N_{a}}(\frac{\Gamma}{2}e^{ik_{a}r_{jl}}+\frac{\gamma}{2}\delta_{jl})\alpha_{l}(t-\frac{r_{jl}}{v_{g}}), 
\end{equation}
with
\begin{equation}
b_{j}(t)=-\frac{i}{2\pi}\sqrt{\frac{\Gamma v_{g}L}{2}}\int_{-\infty }^{\infty }\beta_{k}(0)e^{ikr_{j}-i\delta \omega_{k}t}dk
\end{equation}
being the excitation by the input photon, $\Gamma=2L|g_{k_{a}}|^2/v_{g}$ being the coupling strength between the atom and the guided photon and $r_{jl}=|r_{j}-r_{l}|$. The second term is the collective many-body coupling between the atoms induced by the guided photon modes. It is noted that the collective interactions induced by the guided photon is a long-range effect and there is a time retarded effect for these interactions. The real parts of the collective coupling give the collective damping while the imaginary parts yield an energy shift. Given the initial conditions, we can calculate the atomic excitations at an arbitrary time from Eq. (8). However, the analytical solution for Eq. (8) is usually difficult and we need to resort to numerical methods such as Finite-Difference Time-Domain (FDTD) \cite{FDTD}.

Knowing the atomic excitation $\alpha_{j}(t)$, we can also calculate the amplitude of each guided photon mode at an arbitrary time using Eq. (5). To be specific, if the input photon comes from the left end, the right propagating modes at time $t$ are given by
\begin{equation}
\beta_{\delta k}^{R}(t)=\beta_{k}(0)-i\sqrt{\frac{\Gamma v_{g}}{2L}}\sum_{j=1}^{N_{a}}e^{-i(k_{a}+\delta k)r_{j}}\int_{0}^{t}\alpha_{j}(t')e^{i\delta k v_{g}t'}dt'.
\end{equation}
where $\delta k=k-k_{a}$ with $k>0$. For the left propagating modes we have
\begin{equation}
\beta_{\delta k}^{L}(t)=-i\sqrt{\frac{\Gamma v_{g}}{2L}}\sum_{j=1}^{N_{a}}e^{i(k_{a}+\delta k)r_{j}}\int_{0}^{t}\alpha_{j}(t')e^{i\delta k v_{g}t'}dt'.
\end{equation}
where $\delta k=|k|-k_{a}$ with $k<0$.

To calculate the photon spectrum when $t\rightarrow \infty$, we define
\begin{equation}
\chi _{j}(\delta k)=\int_{-\infty}^{\infty}\alpha_{j}(t)e^{i\delta kv_{g}t}dt
\end{equation}
where $\alpha_{j}(t)=0$ when $t<0$ with $j=1, 2, \cdots, N_{a}$. We then have
\begin{equation}
\alpha_{j}(t)=\frac{v_{g}}{2\pi}\int_{-\infty}^{\infty}\chi _{j}(\delta k)e^{-i\delta kv_{g}t}d\delta k.
\end{equation}
Inserting Eq. (13) into Eq. (8) we obtain a set of linear equations which are given by
\begin{equation}
-i\delta kv_{g}\chi _{j}(\delta k)=b_{j}(\delta k)-\frac{\gamma}{2}\chi_{j}(\delta k)+\sum_{l=1}^{N_{a}}V_{N_{a}}^{jl}(\delta k)\chi _{l}(\delta k)
\end{equation}
where 
\begin{equation}
b_{j}(\delta k)=-i\sqrt{\frac{\Gamma L}{2v_{g}}}\beta_{\delta k}(0)e^{i(k_{a}+\delta k)r_{j}}
\end{equation}
and the collective coupling matrix $V$ is given by
\begin{equation}
V_{N_{a}}(\delta k)=-\frac{\Gamma}{2}\begin{bmatrix}
1 & e^{ika} & \cdots & e^{i(N_{a}-1)ka}\\
e^{ika} & 1 & \cdots & e^{i(N_{a}-2)ka} \\
\vdots   & \vdots & \ddots & \vdots \\
e^{i(N_{a}-1)ka} & e^{i(N_{a}-2)ka} & \cdots & 1
\end{bmatrix}
\end{equation}
with $k=k_{a}+\delta k$.

Let $M(\delta k)=-V_{N_a}(\delta k)+(\gamma/2-i\delta k v_{g})I_{N_a}$ where $I_{N_a}$ is a $N_{a}\times N_{a}$ unit matrix. We have
\begin{equation}
\chi_{j}(\delta k)=\sum_{l=1}^{N_{a}}M^{-1}_{jl}(\delta k)b_{l}(\delta k)
\end{equation}
We can then calculate the photon spectra for the right and left propagating field at $t\rightarrow \infty$:
\begin{eqnarray}
\beta_{\delta k}^{R}(t\rightarrow \infty)&=&\beta_{\delta k}(0)-i\sqrt{\frac{\Gamma v_{g}}{2L}}\sum_{j=1}^{N_{a}}e^{-i(k_{a}+\delta k)r_{j}}\chi _{j}(\delta k), \nonumber \\ \\
\beta_{\delta k}^{L}(t\rightarrow \infty)&=&-i\sqrt{\frac{\Gamma v_{g}}{2L}}\sum_{j=1}^{N_{a}}e^{i(k_{a}+\delta k)r_{j}}\chi _{j}(\delta k). 
\end{eqnarray}

We can calculate the photon pulse shape after the interaction by the Fourier transformations as given by 
\begin{eqnarray}
\beta^{R}_{x}(t)&=&e^{ik_{a}x}\int_{-\infty}^{\infty}\beta^{R}_{\delta k}(t)e^{i\delta k(x-v_{g}t)}d\delta k, \\
\beta^{L}_{x}(t)&=&e^{-ik_{a}x}\int_{-\infty}^{\infty}\beta^{L}_{\delta k}(t)e^{-i\delta k(x+v_{g}t)}d\delta k. 
\end{eqnarray}
We can also calculate the reflectivity $R$ and the transmittivity $T$ of this system and they are given by
\begin{eqnarray}
R&=&\frac{L}{2\pi}\int_{-\infty}^{\infty}|\beta_{\delta k}^{L}(t\rightarrow \infty )|^2 d\delta k, \\
T&=&\frac{L}{2\pi}\int_{-\infty}^{\infty}|\beta_{\delta k}^{R}(t\rightarrow \infty )|^2 d\delta k. 
\end{eqnarray}

In a pure 1D photonic waveguide or a quasi-1D photonic waveguide with $\Gamma\gg \gamma$, we can neglect the spontaneous decay ($\gamma$) to the non-guided modes. In the following sections, we first neglect $\gamma$ and mainly focus on the collective effects induced by the guided modes. We analytically and numerically study the dynamics of a single photon pulse interacting with one atom, two atoms and N equally spaced atoms. In Sec. VI, we briefly discuss the effects of the spontaneous decay to the free space. In all calculations, we assume that the atoms are initially in the ground state and the input single photon pulse has a Gaussian shape which is given by
\begin{equation}
\beta_{\delta k}(0)=\frac{(8\pi)^{1/4}}{\sqrt{\Delta L}}e^{-\frac{(ka+\delta k-k_0)^2}{\Delta^2}},
\end{equation}
where $k_0$ is the wavevector corresponding to the center frequency of the pulse and $\Delta$ is the width in the $k$ space with the full width at half maximum of the spectrum being $\sqrt{2\ln 2}\Delta v_{g}$. A single photon with Gaussian pulse shape can be generated by single photon electric-optic modulation as has been experimentally demonstrated in \cite{Kolchin2008}. We note that $\frac{L}{2\pi}\int_{-\infty }^{\infty }|\beta_{\delta k}(0)|^{2}d\delta k=1$ which is the requirement for a single photon number.

\begin{figure*}
\includegraphics[width=0.6\columnwidth]{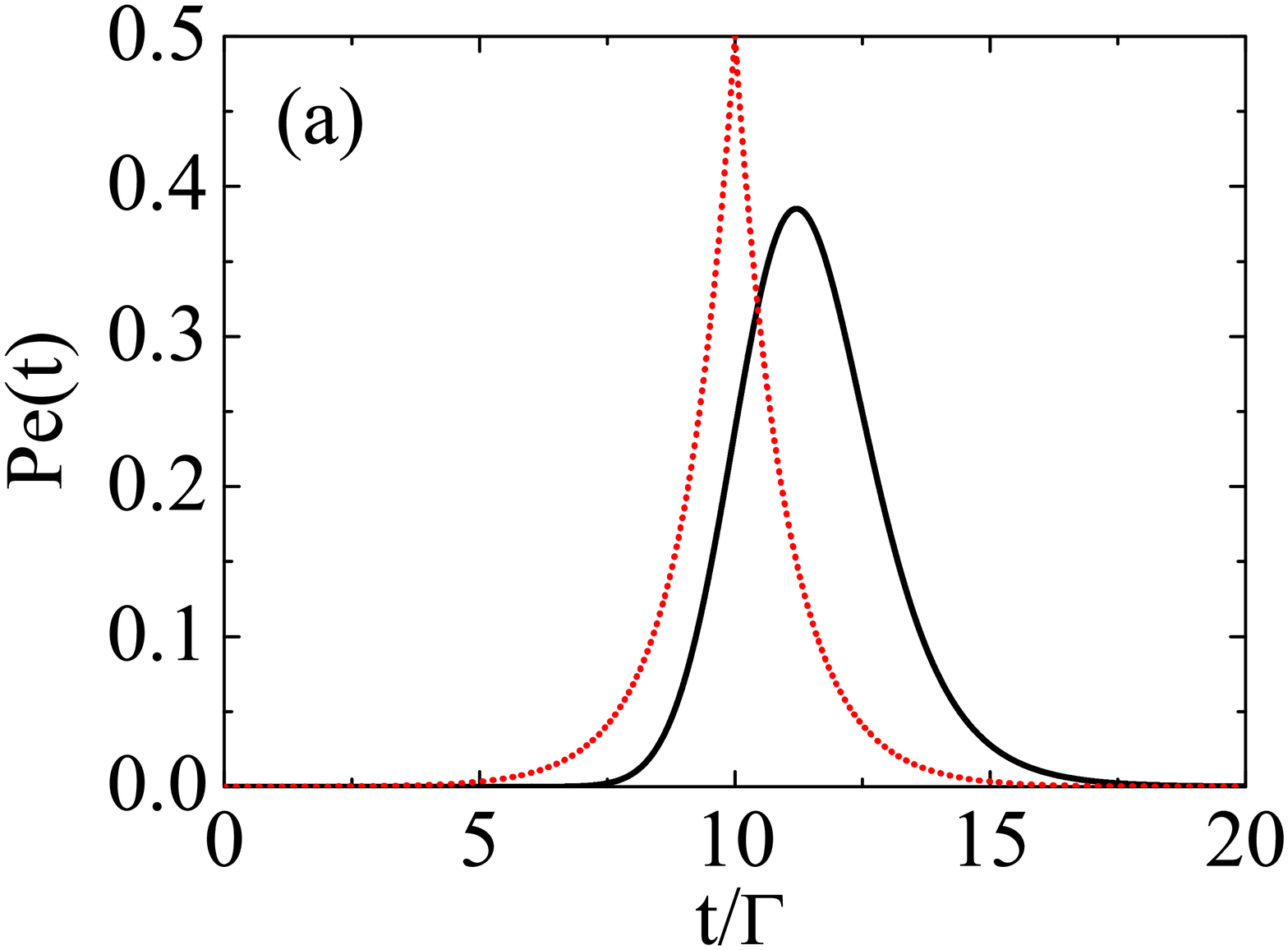}
\includegraphics[width=0.6\columnwidth]{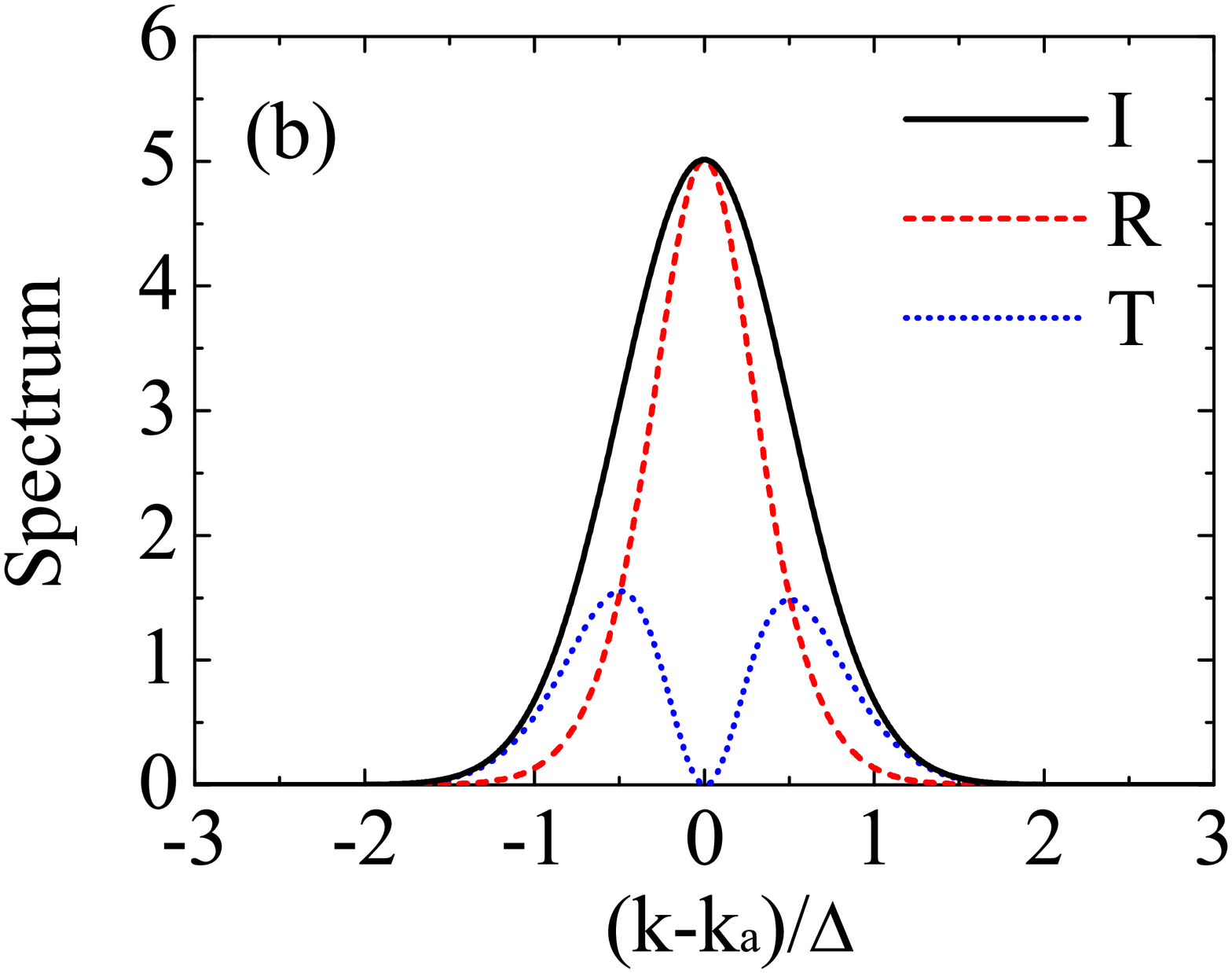}
\includegraphics[width=0.6\columnwidth]{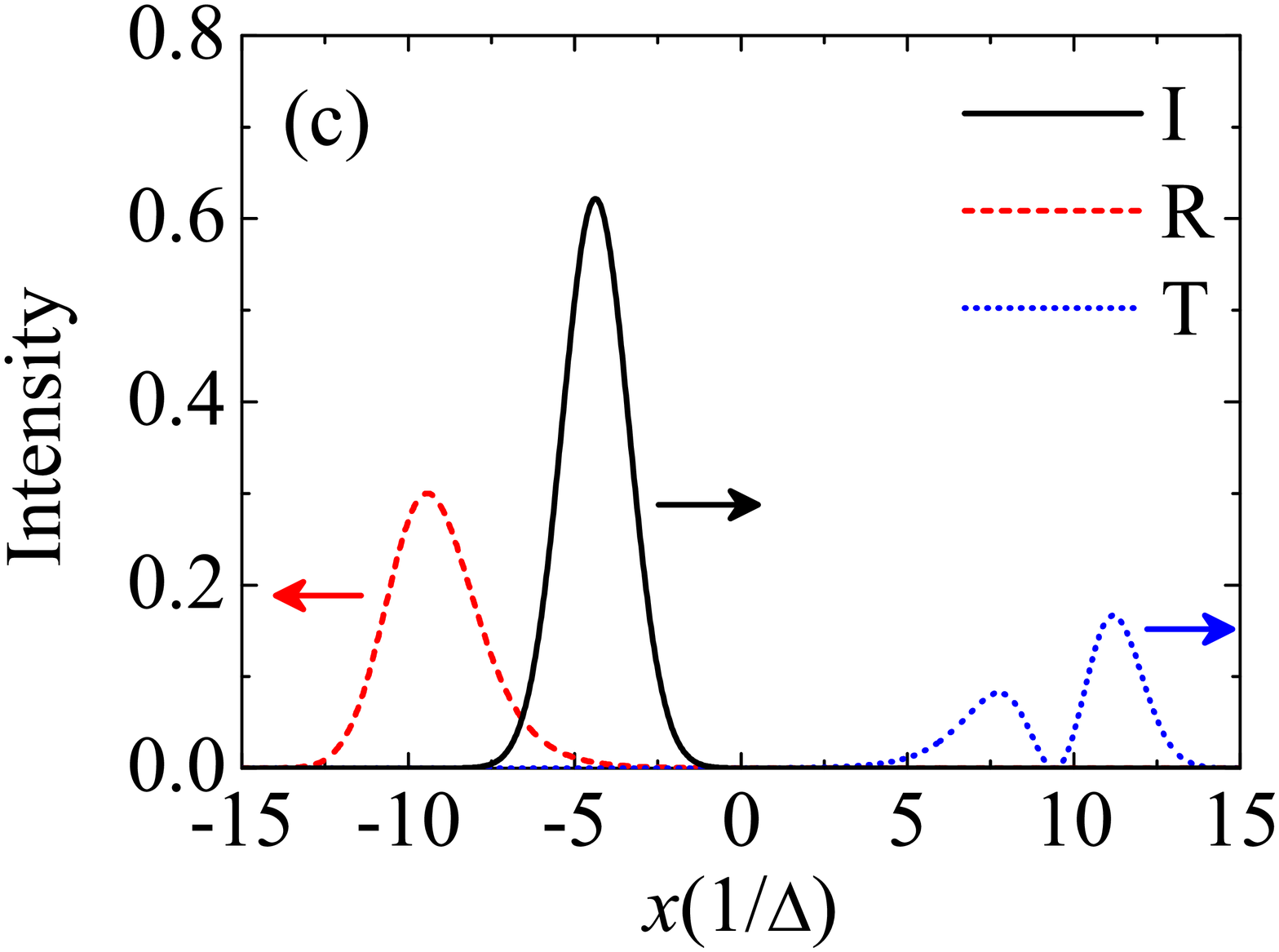}
\includegraphics[width=0.6\columnwidth]{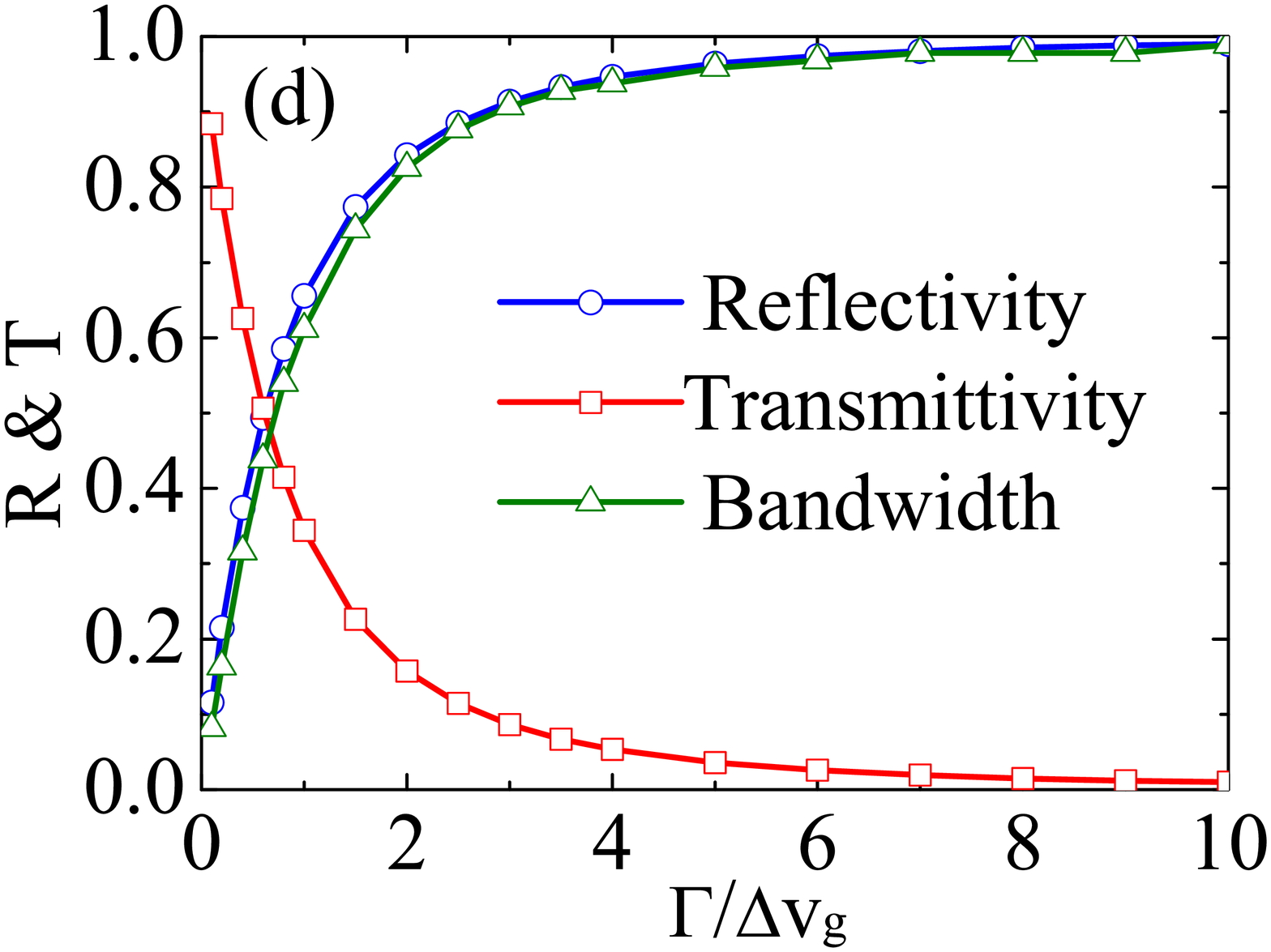}
\includegraphics[width=0.6\columnwidth]{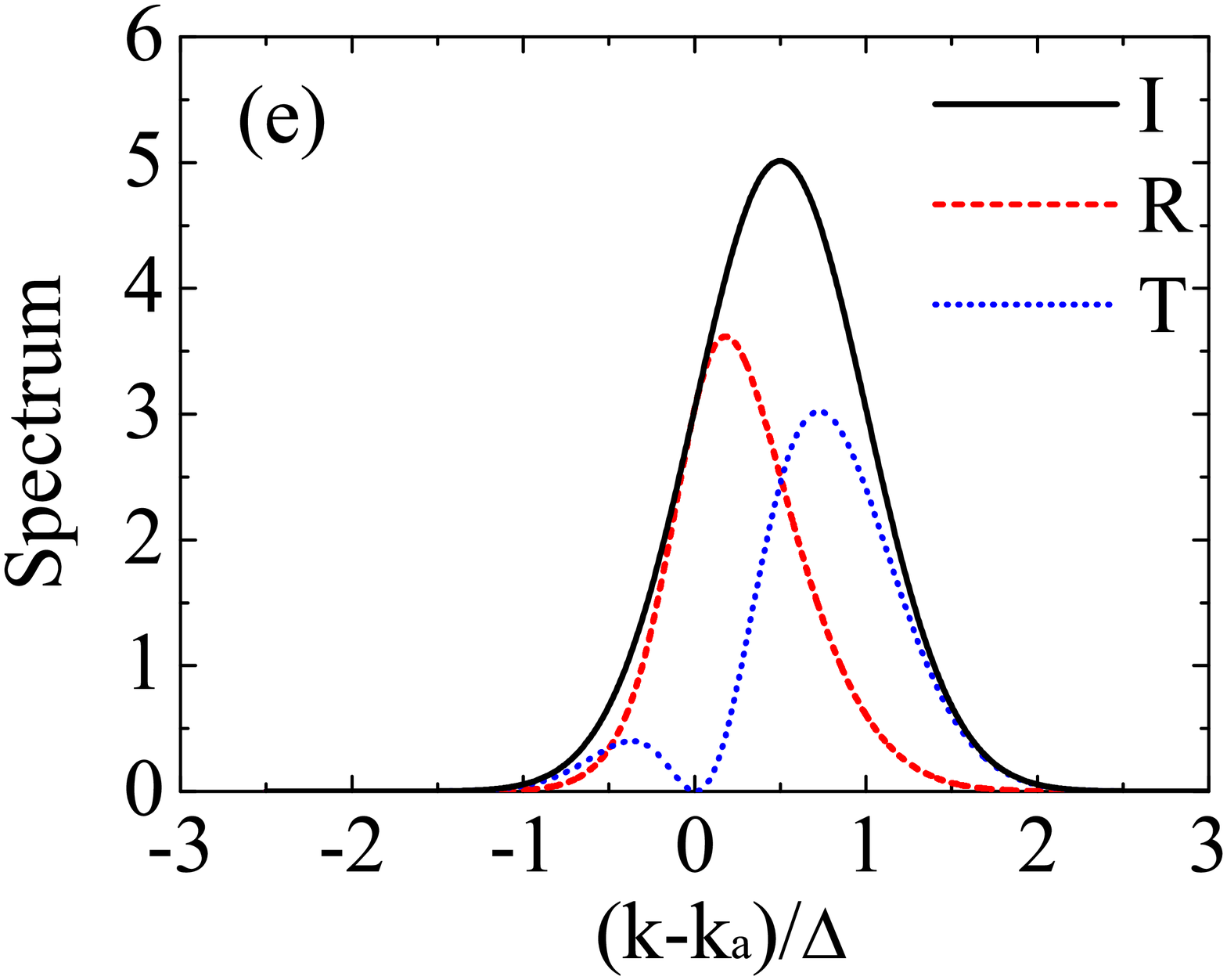}
\includegraphics[width=0.6\columnwidth]{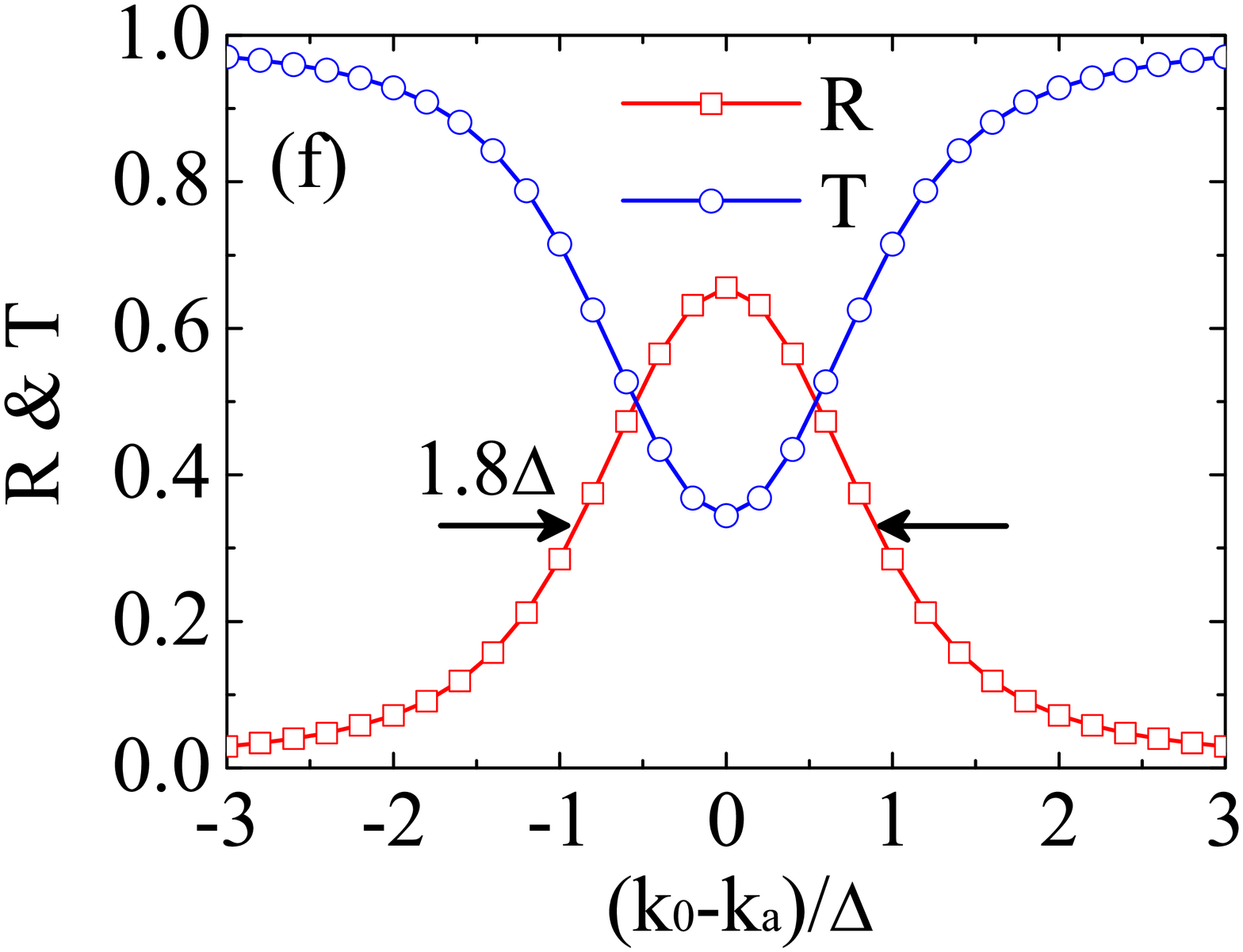}
\caption{(Color online) (a) The atomic excitation as a function time where black solid curve is the result for Gaussian input while the red dotted curve is the result for the inversion pulse given by Ref. \cite{Rephaeli2010}. When $t=0$ the center of the input pulse is  $10/\Delta$ away from the atom. (b) The spectrum of the incoming, reflected and transmitted photon when $\Gamma =\Delta v_g$ and $k_{0}=k_{a}$. (c) The pulse shapes of the incoming (at $t=6/\Gamma$), reflected (at $t=20/\Gamma$) and transmitted (at $t=20/\Gamma$) photon when $\Gamma =\Delta v_g$ and $k_{0}=k_{a}$. The atom is at the position $x=0$.  (d) The reflectivity, transmittivity and relative full width at half maximum of the reflected pulse as a function of coupling strength. The unit of the reflection bandwidth is $\sqrt{2\ln 2}\Delta v_{g}$. (e) The spectrum of the incoming, reflected and transmitted photon when $\Gamma =\Delta v_{g}$ and $k_{0}-k_{a}=\Delta/2$. (f) The reflectivity and transmittivity as a function of detuning $k_{0}-k_{a}$ when $\Gamma =\Delta v_g$. Here, I is incoming; R is reflection; T is transmission. }
\end{figure*}

\section{Single Atom}

In \cite{Chen2011}, Chen {\it et al.} have numerically studied the case of a single Gaussian photon pulse scattered by a single atom coupled to a 1D waveguide. Here we present some analytical results based on the method shown in the previous section. 

From Eq. (8), the atomic excitation for one-atom case with $\gamma=0$ is governed by 
\begin{equation}
\dot{\alpha}_{1}(t)=b_{1}(t)-\frac{\Gamma}{2}\alpha_{1}(t),
\end{equation}
where $b_{1}(t)$ is given by Eq. (9) with $j=1$. The formal solution of this equation is given by
\begin{equation}
\alpha (t)=-iS\{\mbox{erf}(C+\sqrt{B}t_0)-\mbox{erf}(C)\}e^{-\frac{\Gamma}{2}t}
\end{equation}
where $C=(A-2Bt_0)/2\sqrt{B}$ with $t_0=r_1/v_{g}$, $A=-\Gamma/2+i(k_{0}-k_{a})v_{g}$, and $B=\Delta^{2}v_{g}^{2}/4$. $S=(\pi/8)^{1/4}\sqrt{\Gamma/\Delta v_{g}}e^{-At_0+A^2/4B}e^{ik_a r_1}$, and $\mbox{erf}(x)=(2/\sqrt{\pi})\int_{0}^{x}e^{-t^2}dt$ is the error function. This result has been derived in Ref. \cite{Chen2011} with $k_{0}=k_{a}$. Here we generalize this result to include the case when $k_{0}\neq k_{a}$, i.e., when the center frequency of the input photon pulse is not equal to the transition frequency of the atom.

The atomic excitation as a function of time when $\Gamma=\Delta v_{g}$ and $k_{0}=k_{a}$ is shown as the black solid line in Fig. 2(a). When $t=0$, the center of the input photon pulse is $10/\Delta$ away from the atom. The atom is first excited and then deexcited when the photon pulse propagates through the atomic chain. The photon pulse incident from one direction can be separated as odd and even modes \cite{Rephaeli2010}. Since only even mode can couple to the atom, the maximum excitation by a photon incident from one direction is $50\%$. The maximum atomic excitation for the Gaussian input in our example is about $40\%$ which is close to the $50\%$ limit. The $50\%$ limit can be reached by the inversion pulse incident from one side as discussed in Ref. \cite{Rephaeli2010} (see the red dotted line in Fig. 2(a)). 

For one atom,  Eq. (14) becomes
\begin{equation}
-i\delta kv_{g}\chi _{1}(\delta k)=b_{1}(\delta k)-\frac{\Gamma}{2}\chi _{1}(\delta k)
\end{equation}
whose solution is
\begin{equation}
\chi _{1}(\delta k)=-i\sqrt{\frac{\Gamma L}{2v_{g}}}e^{i(k_{a}+\delta k)r_{1}}\frac{\beta_{\delta k}(0)}{\frac{\Gamma}{2}-i\delta kv_{g}}.
\end{equation}
The photon spectra after the interaction can then be calculated from Eqs. (18) and (19), and given by
\begin{eqnarray}
\beta_{\delta k}^{R}(t\rightarrow \infty)&=&-\frac{(2i\delta kv_{g}/\Gamma)\beta_{\delta k}(0)}{1-2i\delta kv_{g}/\Gamma}, \\
\beta_{\delta k}^{L}(t\rightarrow \infty)&=&-e^{2i(k_{a}+\delta k)r_{1}}\frac{\beta_{\delta k}(0)}{1-2i\delta kv_{g}/\Gamma}. 
\end{eqnarray}
with $\beta_{\delta k}(0)$ given by Eq. (24).
The spectra of the photon before and after the interaction when $k_{0}=k_{a}$ and $\Gamma=\Delta v_{g}$ are shown in Fig. 2(b) where we see that the reflected spectrum is similar to the initial spectrum but the transmitted spectrum has two peaks.  However, the reflected spectrum is narrower than the incoming spectrum and the reflected bandwidth can be controlled by changing the coupling strength. The smaller the coupling strength becomes, the narrower the reflected bandwidth is. In this case, there is about $66\%$ possibility that the photon is reflected. When $\delta k=0$, we have $|\beta_{\delta k}^{L}(t\rightarrow \infty)|^{2}=|\beta_{\delta k}(0)|^2$ and $|\beta_{\delta k}^{R}(t\rightarrow \infty)|^{2}=0$ which means that the resonant frequency component is completely reflected and it is independent of the input pulse shape. This is consistent with the stationary calculations in Ref. \cite{Shen2005a}. The positions of the peaks $\delta k_{\pm}$ in the transmitted spectrum are given by
\begin{equation}
\frac{\delta k_{\pm}}{\Delta}=\pm \frac{\eta }{2\sqrt{2}}\left [\left (-1 +\sqrt{1+\frac{8}{\eta ^2}}\right ) \right ]^{1/2}
\end{equation}
where $\eta=\Gamma/\Delta v_{g}$ is the the ratio between the coupling strength and the input photon bandwidth. We can see that the peaks depends only on this ratio. When $\eta\rightarrow 0$, $\delta k_{\pm}\rightarrow 0$. The splitting of these two peaks increases when we increase $\eta $ and it approaches $\sqrt{2}\Delta$ when $\eta \rightarrow \infty $. When $\eta =1$, $\delta k_{\pm}=\pm \frac{1}{2}\Delta$ which are the peak positions of the transmitted spectrum shown in Fig. 2(b).

The photon pulse shape before and after the interaction can be calculated by Eqs. (19) and (20) and they are shown in Fig. 2(c) where the black solid curve is the incoming photon pulse while the red dotted curve is the reflected photon pulse and the blue dashed curve is the transmitted photon pulse. The reflected pulse shape is similar to the incoming pulse shape while the transmitted pulse shape is quite different where there are two peaks due to the interference between the incoming photon amplitude and the reemitted photon amplitude. 

The reflectivity and transmittivity of this system, obtained from Eqs. (21) and (22), are given by
\begin{eqnarray}
R&=&\sqrt{\frac{\pi}{2}}\int_{-\infty }^{\infty}\frac{e^{-2y^{2}}}{1+4y^2/\eta ^{2}}dy, \\
T&=&1-R.
\end{eqnarray}
The reflectivity and transmittivity of a single atom system also depends only on the ratio $\Gamma/\Delta v_{g}$. When $\eta \rightarrow 0$, $R\rightarrow 0$ and $T\rightarrow 1$, i.e, the photon can completely transmit when there is no coupling. When $\eta \rightarrow \infty $, $R\rightarrow 1$ and $T\rightarrow 0$, i.e, the photon is completely reflected when the coupling strength is very large comparing with the photon bandwidth. The reflectivity and transmittivity, as a function of the ratio $\Gamma/\Delta v_{g}$, are shown in Fig. 2(d) where we see that the reflectivity increases when $\Gamma/\Delta v_{g}$ increases and it can approach $100\%$ reflectivity when $\Gamma/\Delta v_{g}\gg 1$. The reflection bandwidth can also be controlled by changing the ratio $\Gamma/\Delta v_{g}$. The smaller the $\Gamma/\Delta v_{g}$ is, the narrower the reflection bandwidth is (see the green line with triangle symbol in Fig. 2(d)). This may be used to generate a single photon with ultra-narrow linewidth.

We also study the case when the center frequency of the incoming photon is detuned from the resonant frequency, i.e., $k_{0}\neq k_{a}$. For example, the result when $k_{0}-k_{a}=\Delta/2$ is shown in Fig. 2(e) where we can see that the spectrum is quit different. Both the reflected and transmitted spectra are asymmetric. However, the resonant frequency component here is also completely reflected. The reflectivity and transmittvity, as a function of center frequency detuning $k_{0}-k_{a}$ when $\Gamma=\Delta v_{g}$, are shown in Fig. 2(f) where we see that the reflectivity decreases as the detuning increases and the highest reflectivity here is about $66\%$. The full width at half maximum of the reflection curve is about $1.8\Delta v_{g}$.

\section{Two Atom}

In this section, we study how a single photon pulse is scattered by a two-atom system coupled to a 1D waveguide. This is the simplest collective many-body system.

From Eq. (8), the atomic excitation dynamics for two-atom system are given by 
\begin{equation}
\dot{\vec{\alpha}}(t)=\vec{b}(t)+V_{2}(\delta k=0)\vec{\alpha}(t-\frac{a}{v_{g}}), 
\end{equation}
where $\vec{\alpha}=(\alpha_{1}, \alpha_{2})^T$, $\vec{b}=(b_{1}, b_{2})^T$, and the effective collective coupling matrix is
\begin{equation}
V_{2}(\delta k=0)=-\frac{\Gamma}{2}\begin{bmatrix}
1 & e^{ik_{a}a} \\
e^{ik_{a}a} & 1
\end{bmatrix}
\end{equation}
with $a$ being the atomic separation. 
The analytical solutions for the coupled equations (33) are difficult to obtain.  
We can apply numerical method to calculate the atomic excitation in arbitrary time. 

To calculate the spectrum, we follow the procedures shown in Sec. II. For two-atom system, Eq. (14) becomes
\begin{equation}
-i\delta kv_{g}\vec{\chi}(\delta k)=\vec{b}(\delta k)+V_{2}(\delta k)\vec{\chi}(\delta k)
\end{equation}
where $\vec{\chi}=(\chi_{1}, \chi_{2})^T$, $\vec{b}=(b_{1}, b_{2})^T$, and $V_{2}(\delta k)$ is given by Eq. (16) with $N_{a}=2$. The solutions for these equations can be readily obtained. After inserting these solutions into Eqs. (18) and (19), we can get the photon spectra when $t\rightarrow \infty $ which are given by
\begin{eqnarray}
\beta_{\delta k}^{R}&=&\beta_{\delta k}(0)\frac{-4\delta k^2v_{g}^{2}/\Gamma^{2}}{(1-2i\delta kv_{g}/\Gamma)^2-e^{2ika}}, \\
\beta_{\delta k}^{L}&=&\beta_{\delta k}(0)e^{2ikr_{1}}\frac{(1+e^{2ika})(1-2i\delta kv_{g}/\Gamma)-2e^{2ika}}{(1-2i\delta kv_{g}/\Gamma)^2-e^{2ika}}, \nonumber \\
\end{eqnarray} 
where $k=k_{a}+\delta k$.

Before discussing the detailed results, we first look at what has been modified by the effective collective coupling. Since $\alpha_{j}(t)$ are the one-photon excitation amplitudes, we have $\vec{\alpha}(t)=\alpha_{1}(t)|eg\rangle +\alpha_{2}(t)|ge\rangle$ and the coupling matrix can be written as $V_{2}(\delta k=0)=-(\Gamma/2)(|eg\rangle\langle eg|+|ge\rangle\langle ge|+e^{ik_{a}a}|eg\rangle\langle ge|+e^{ik_{a}a}|ge\rangle\langle eg|)$. 
It is noted that this coupling matrix is non-Hermitian because it includes the effects of dissipation to the photon modes \cite{Rau2002}. The eigenvalues and the eigenstates of $V_{2}(\delta k=0)$ can be calculated to be
\begin{eqnarray}
\lambda_{\pm}&=&-\frac{\Gamma}{2}(1\pm\cos(k_{a}a))\pm i\frac{\Gamma}{2}\sin(k_{a}a), \\
|\pm\rangle &=& \frac{1}{\sqrt{2}}(|eg\rangle \pm |ge\rangle).
\end{eqnarray}
The real parts of the eigenvalues give the collective coupling strength while the imaginary parts give the energy shifts. The eigenstates of the coupling matrix are two Dicke states. We can change the atomic separation to tune the collective couping strength and the energy shift. In the following we compare the results for four different atomic separations, i.e., $a=\lambda/4, \lambda/2, \lambda/8$ and $3\lambda/8$. 

\begin{figure*}
\includegraphics[width=0.6\columnwidth]{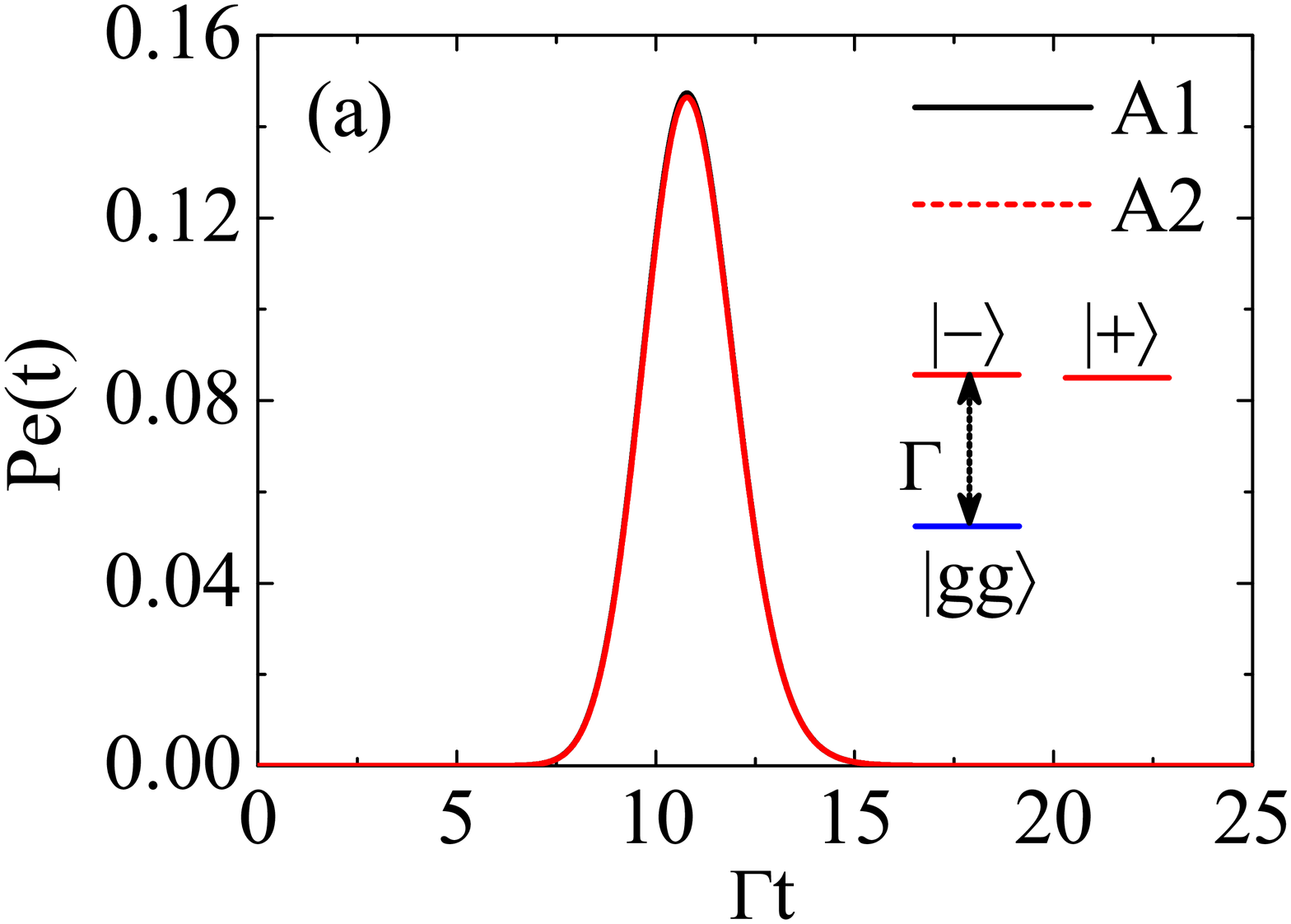}
\includegraphics[width=0.6\columnwidth]{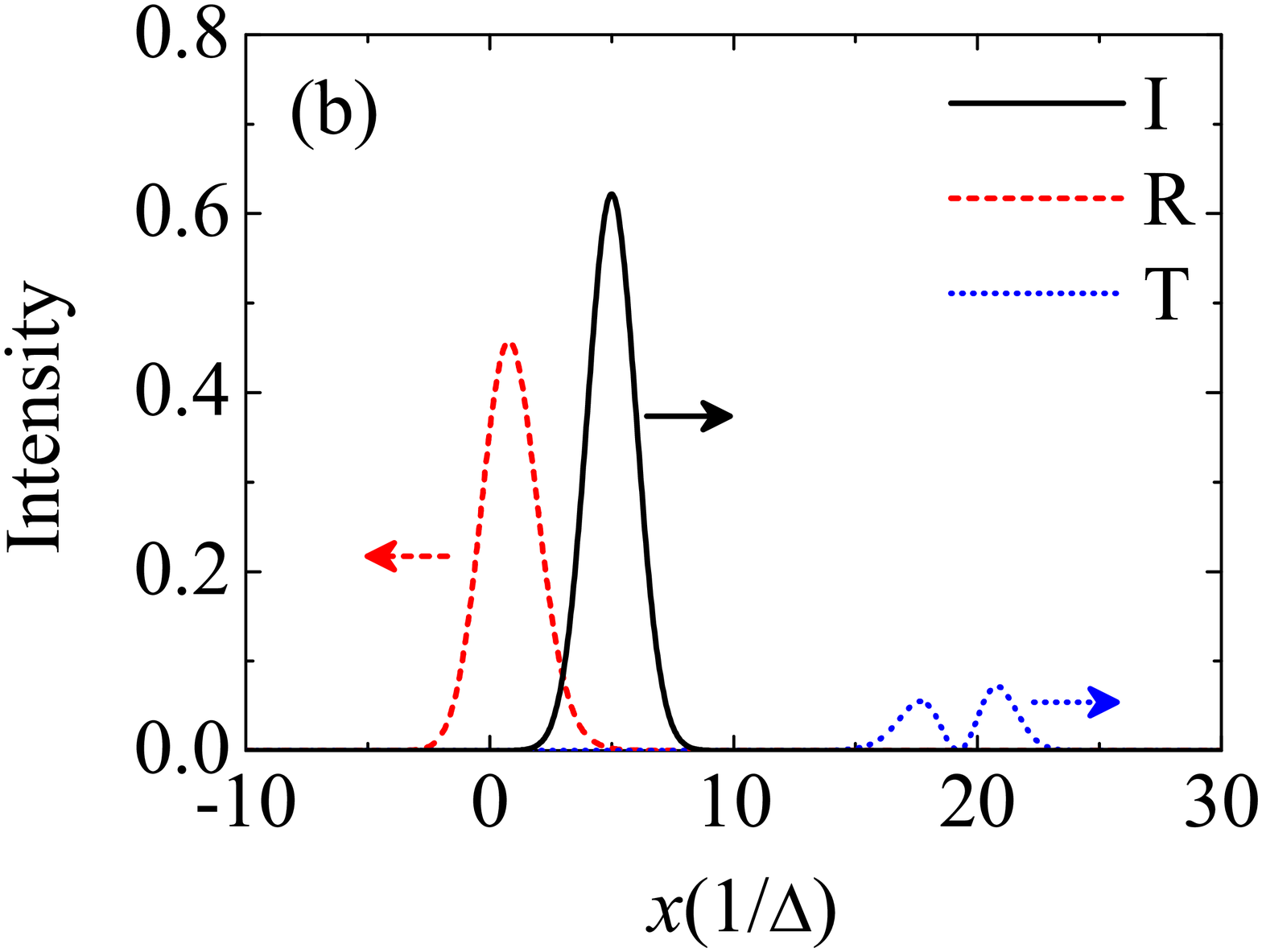}
\includegraphics[width=0.6\columnwidth]{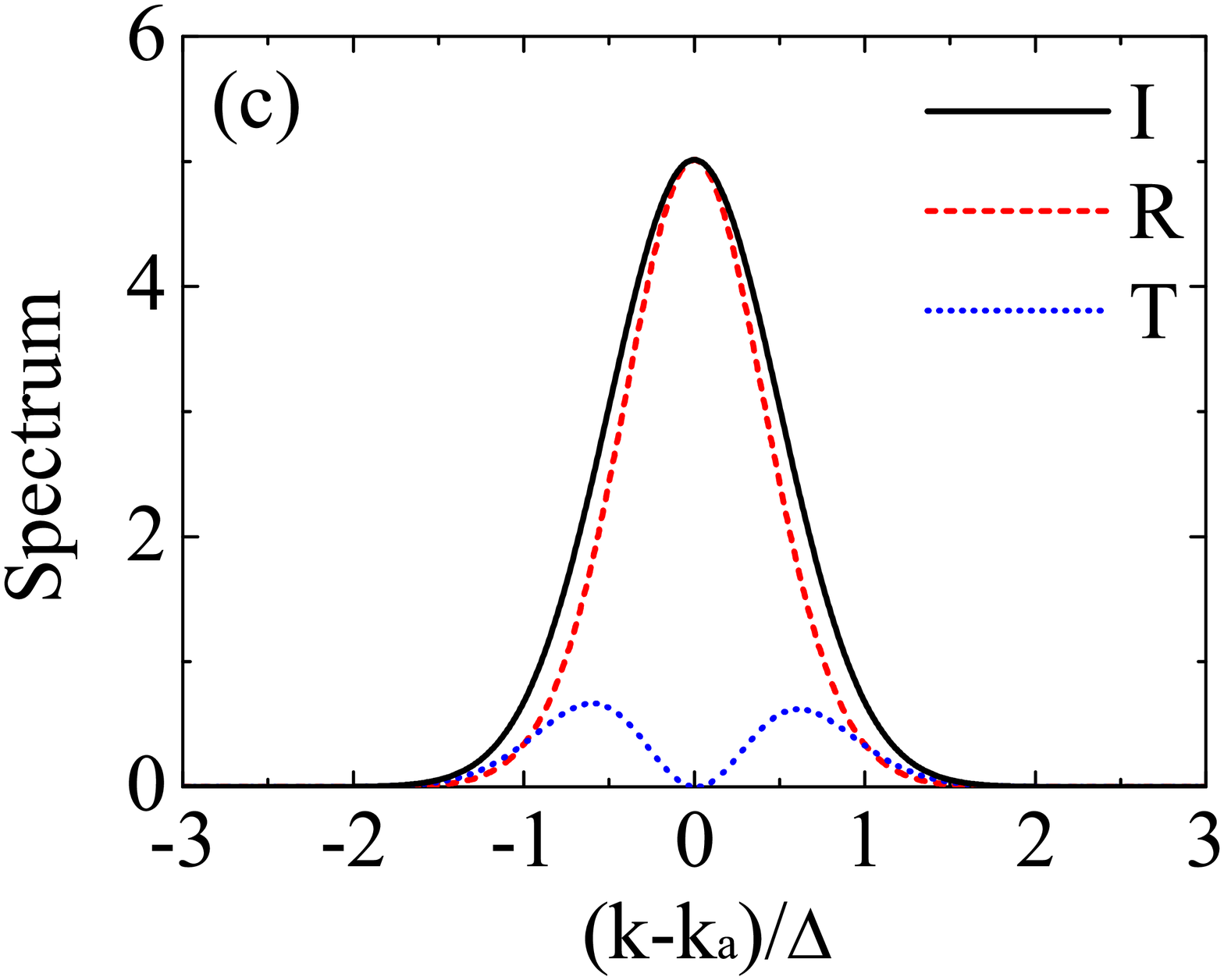}
\includegraphics[width=0.6\columnwidth]{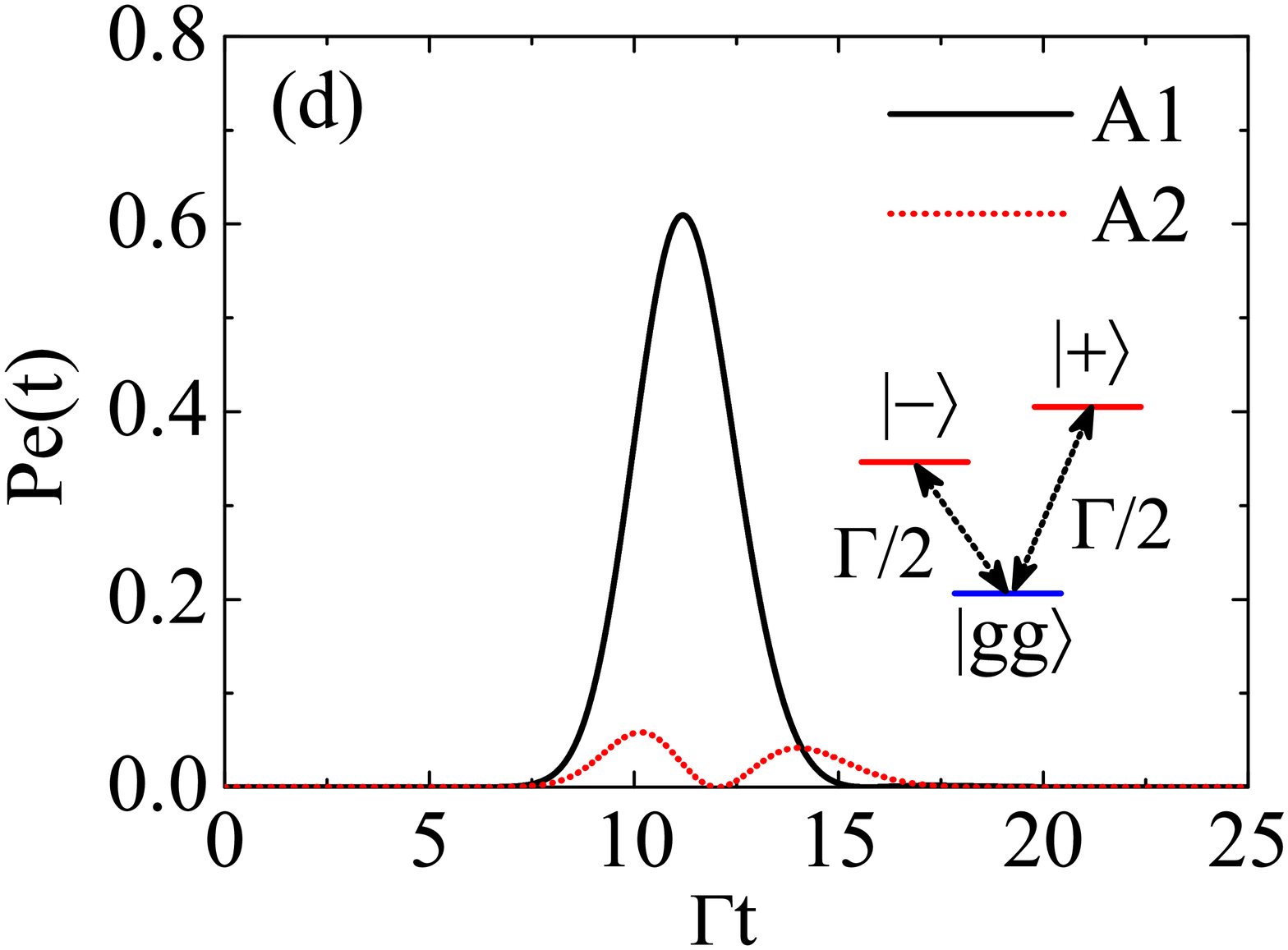}
\includegraphics[width=0.6\columnwidth]{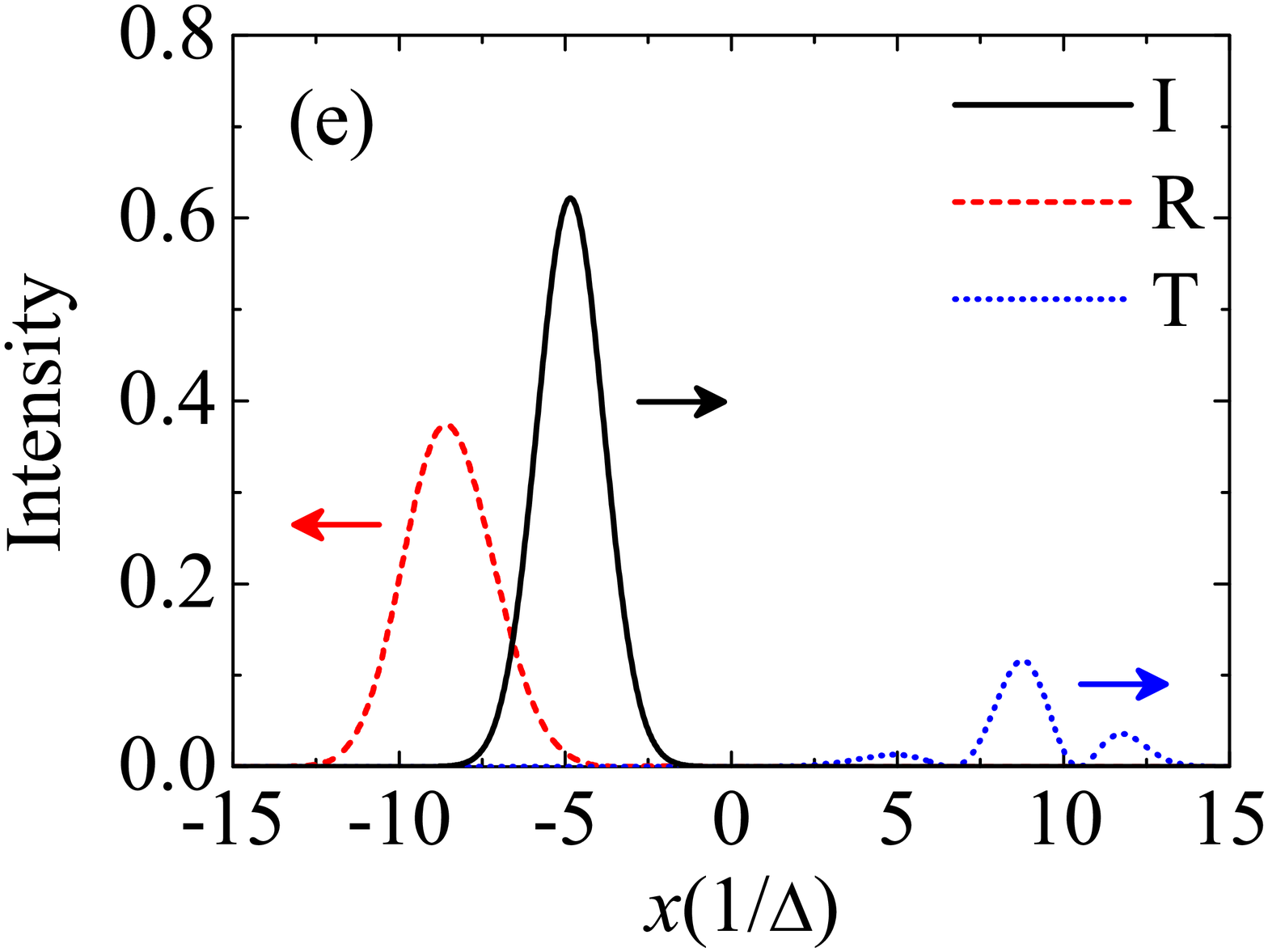}
\includegraphics[width=0.6\columnwidth]{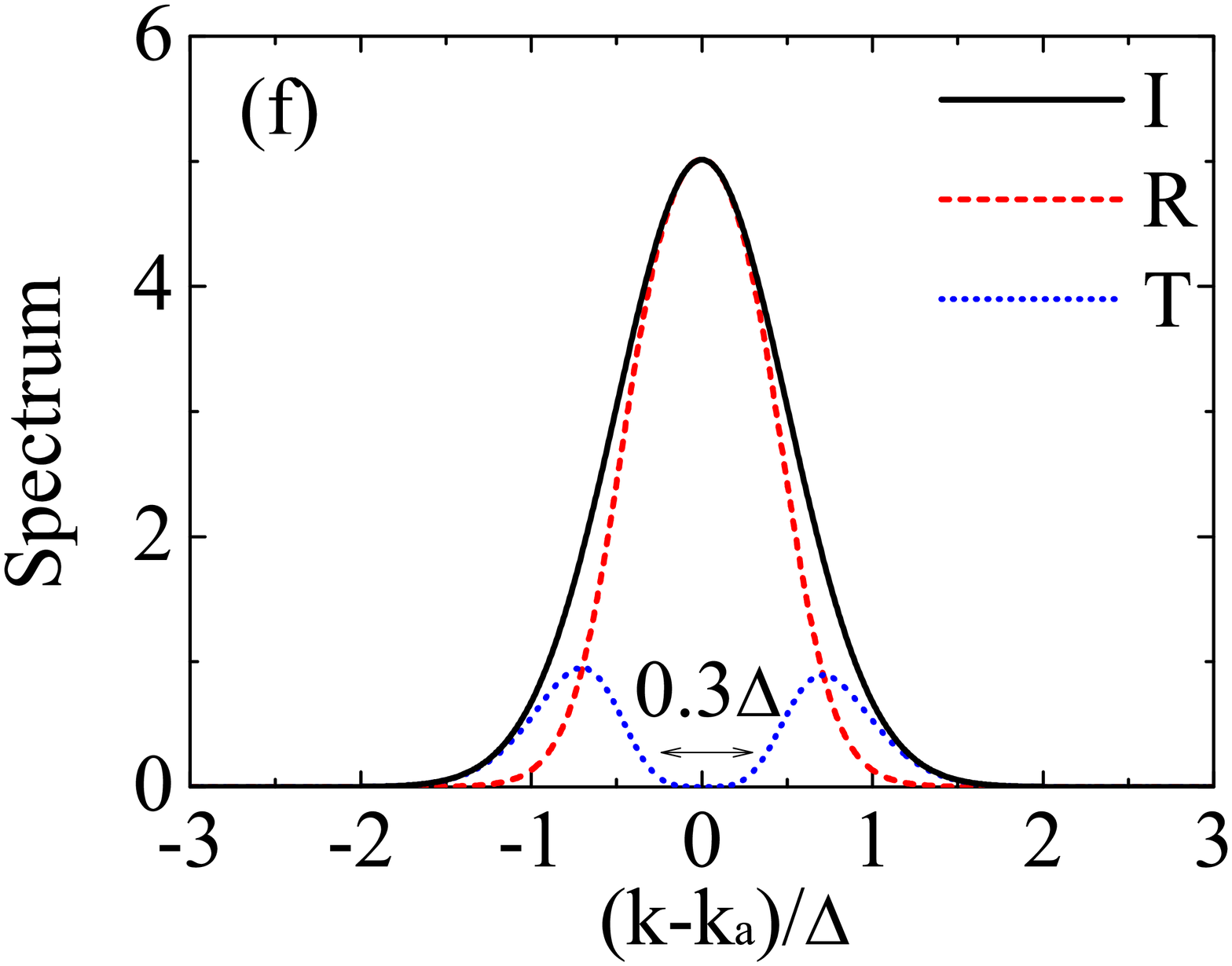}
\includegraphics[width=0.6\columnwidth]{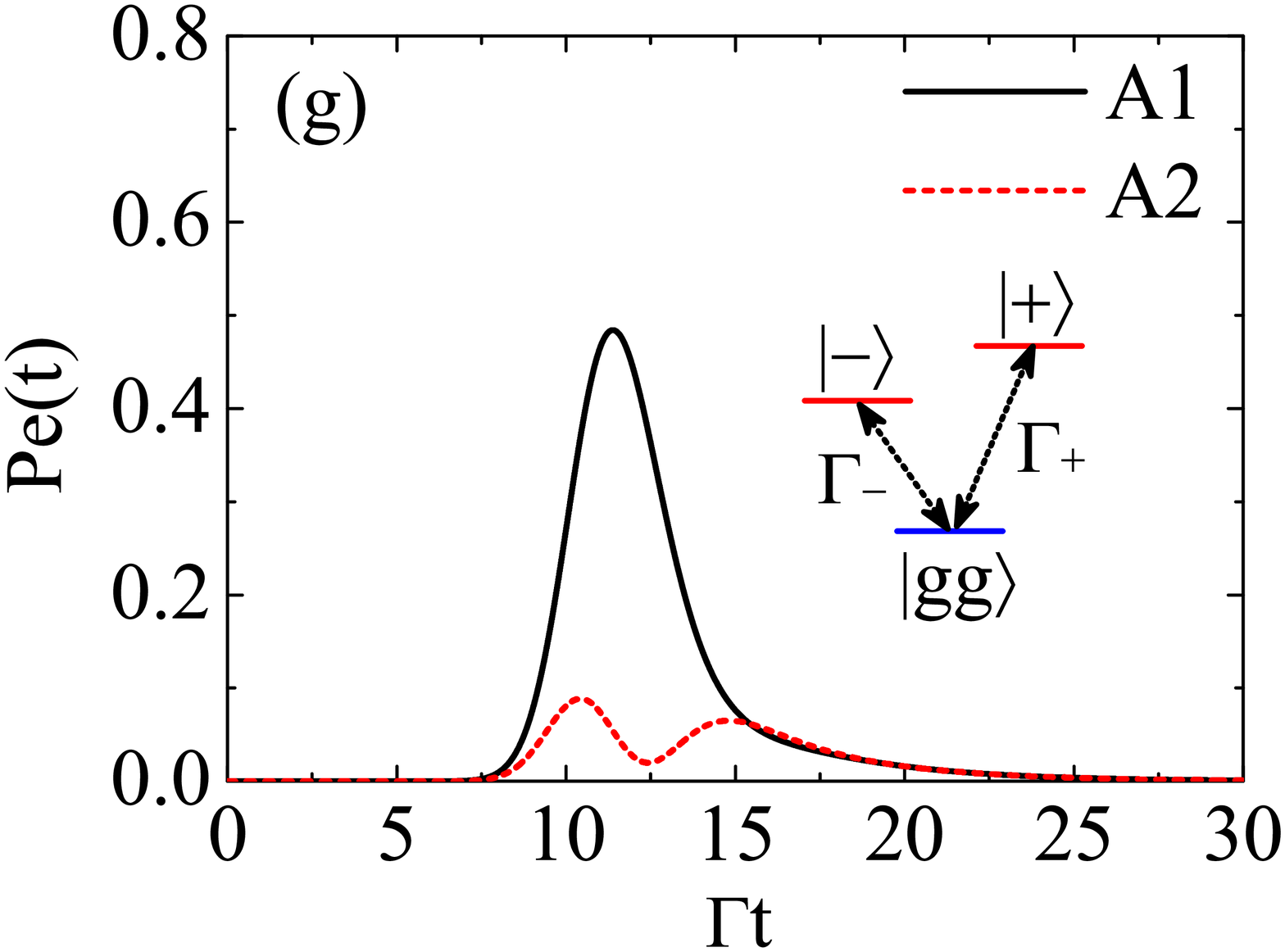}
\includegraphics[width=0.6\columnwidth]{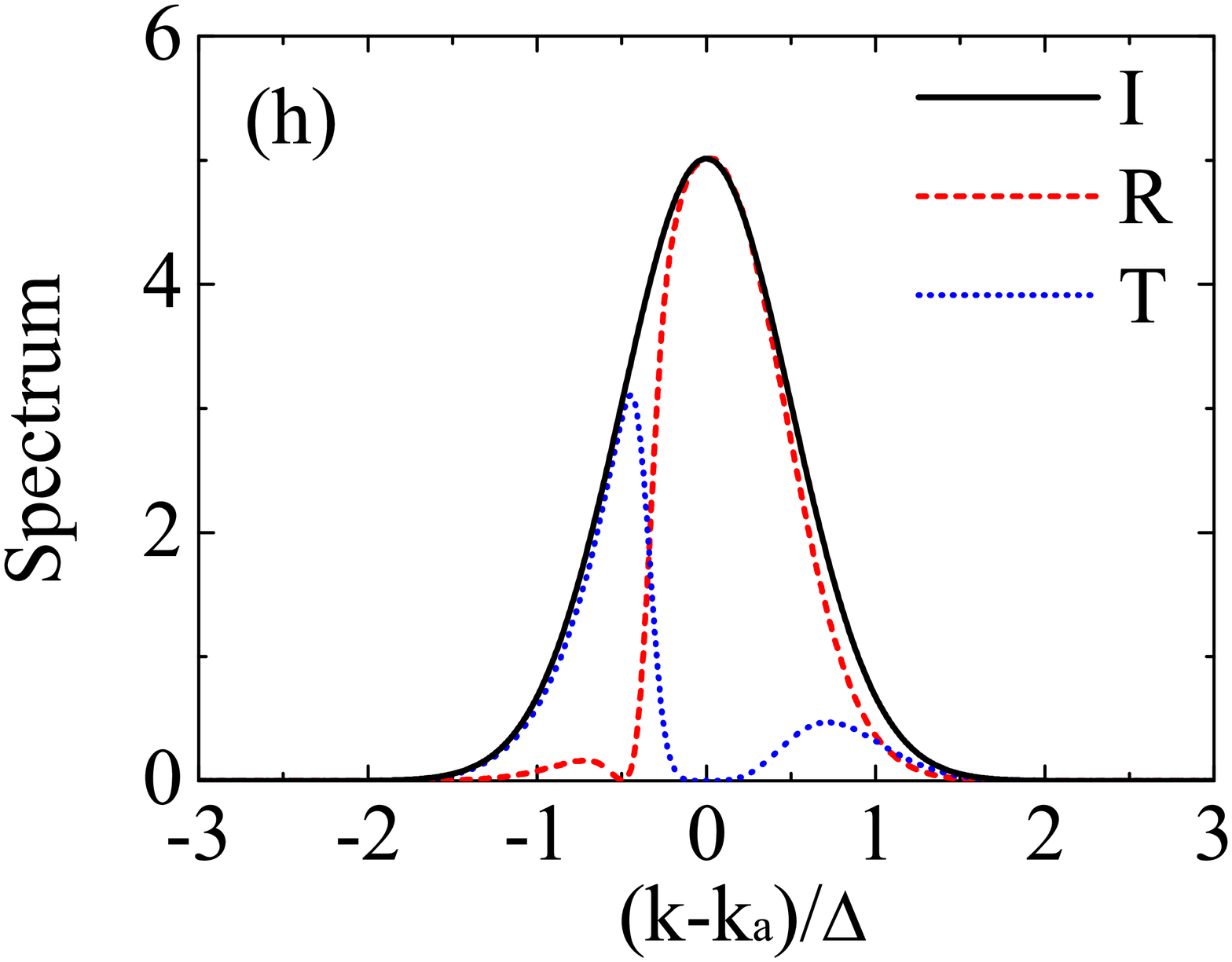}
\includegraphics[width=0.6\columnwidth]{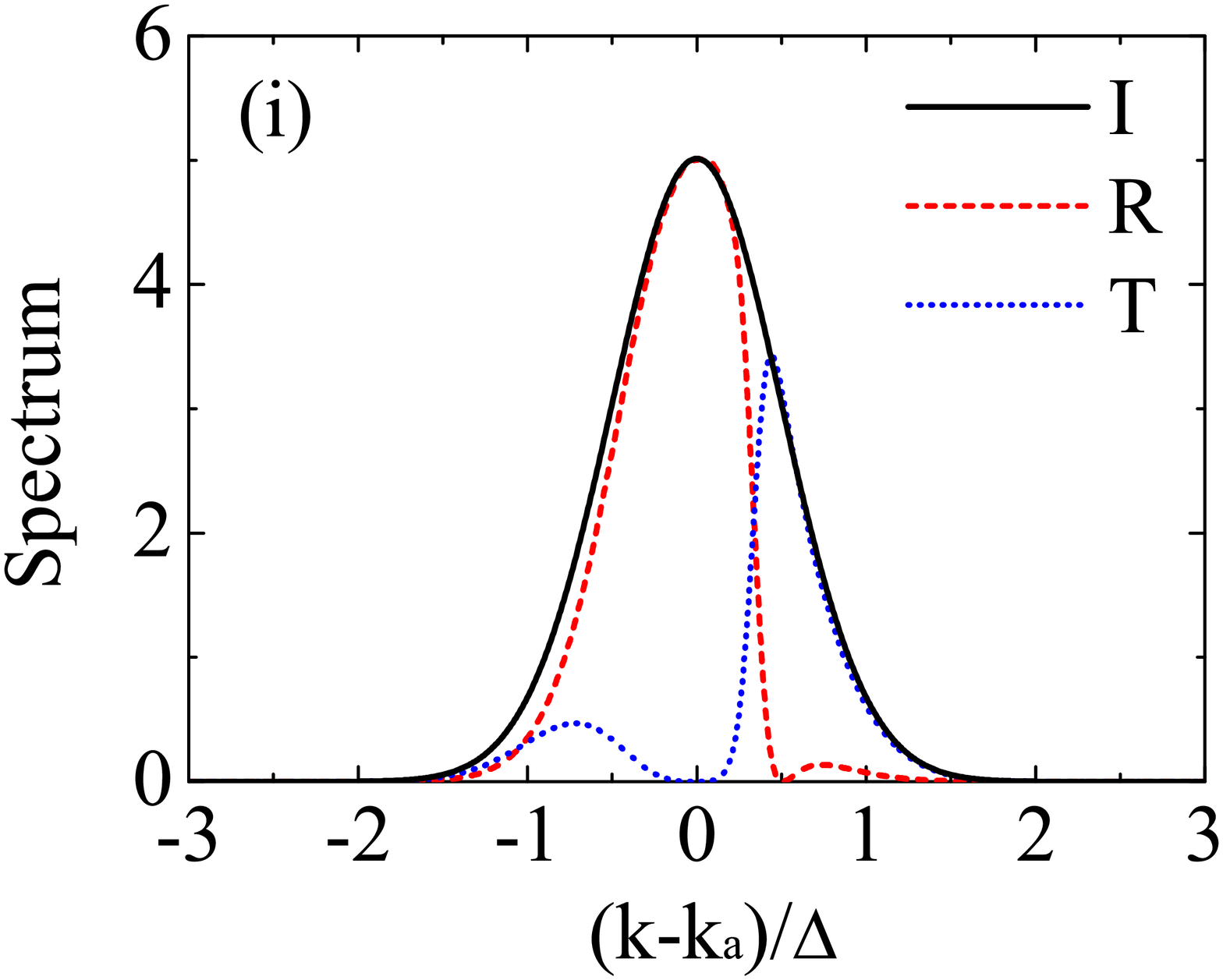}
\caption{(Color online) (a, d, g) The atomic excitations (A1: atom 1 and A2: atom 2) as a function of time. When $t=0$ the center of the input pulse is  $10/\Delta$ away from the first atom. Here, A1 is for atom 1 and A2 is for atom 2. (b, e) The pulse shapes of the incoming (I, at $t=6/\Gamma$), reflected (R, at $t=20/\Gamma$) and transmitted (T, at $t=20/\Gamma$) photon. (c, f, h, i) The spectrum (arb. units) of the incoming (I), reflected (R) and transmitted (T) photon. Parameters: $\Gamma =\Delta v_g$, (a-c) $a=\lambda/2$, (d-f) $a=\lambda/4$, (g, h) $a=\lambda/8$, (i) $a=3\lambda/8$. }
\end{figure*}

The results are shown in Fig. 3 where we assume that the center of the input pulse is initially $10/\Delta$ away from the first atom. In Fig. 3(a-c) we present the results when $a=\lambda/2$ and $\Gamma=\Delta v_{g}$. In this case, the eigenvalues of the coupling matrix are $\lambda_{+}=0$ and $\lambda_{-}=-\Gamma$. The collective coupling between the two atoms when $a=\lambda/2$ modifies the collective couping strength but does not shift the energy. The two single-photon excitation eigenstates are degenerate. Different from the usual case, here $|\lambda_{-}\rangle$ is superradiant while $|\lambda_{+}\rangle$ does not couple to the waveguide modes. This is because of the extra phase factor $e^{i\pi}$ caused by the spatial separation between the two atoms. Therefore, the system is prepared in the $|\lambda_{-}\rangle$ eigenmode where the two atoms can have almost equal probabilities to be excited (Fig. 3(a)). Since only one eigenmode couples to the ground state, the photon pulse shape (Fig. 3(b)) and photon spectrum (Fig. 3(c)) before and after the interaction are similar to the one-atom cases. However, the reflectivity here is about $84\%$ which is larger than the one-atom case due to the enhanced collective coupling strength.

For $a=\lambda/4$, the eigenvalues of the coupling matrix are $\lambda_{\pm}=-\Gamma/2\pm i\Gamma/2$. The collective interaction only cause energy shift but does not modify the collective coupling strength. The two eigenmodes have different energies but they couple to the ground state with the same coupling strength. In Fig. 3(d) we show the excitation probabilities of the two atoms as a function of time where the black solid line is the excitation for the first atom while the red dotted line is the excitation for the second atom. Due to the interference between the two excitation channels, the first atom has a maximum excitation probability larger than $50\%$ which is not allowed for single atom case. The second atom has much smaller excitation probability and experiences two cycles of excitation and deexcitation and reaches a minimum excitation in between due to the destructive interference. The photon pulses before and after the scattering are shown in Fig. 3(e) where we can see that the reflected pulse is similar to incoming pulse while the transmitted pulse has three peaks instead of two peaks as in the one-atom scattering. The spectrum is shown in Fig. 3(f). Different from the single atom case, there is a frequency bandwidth around the resonant frequency that can not be transmitted. The bandwidth where less than $1\%$ transmittance is about $30\%$ of input pulse width. This is similar to the photonic bandgap effects. In principle, we can tune the coupling strength to tune this bandwidth.

\begin{figure*}
\includegraphics[width=0.6\columnwidth]{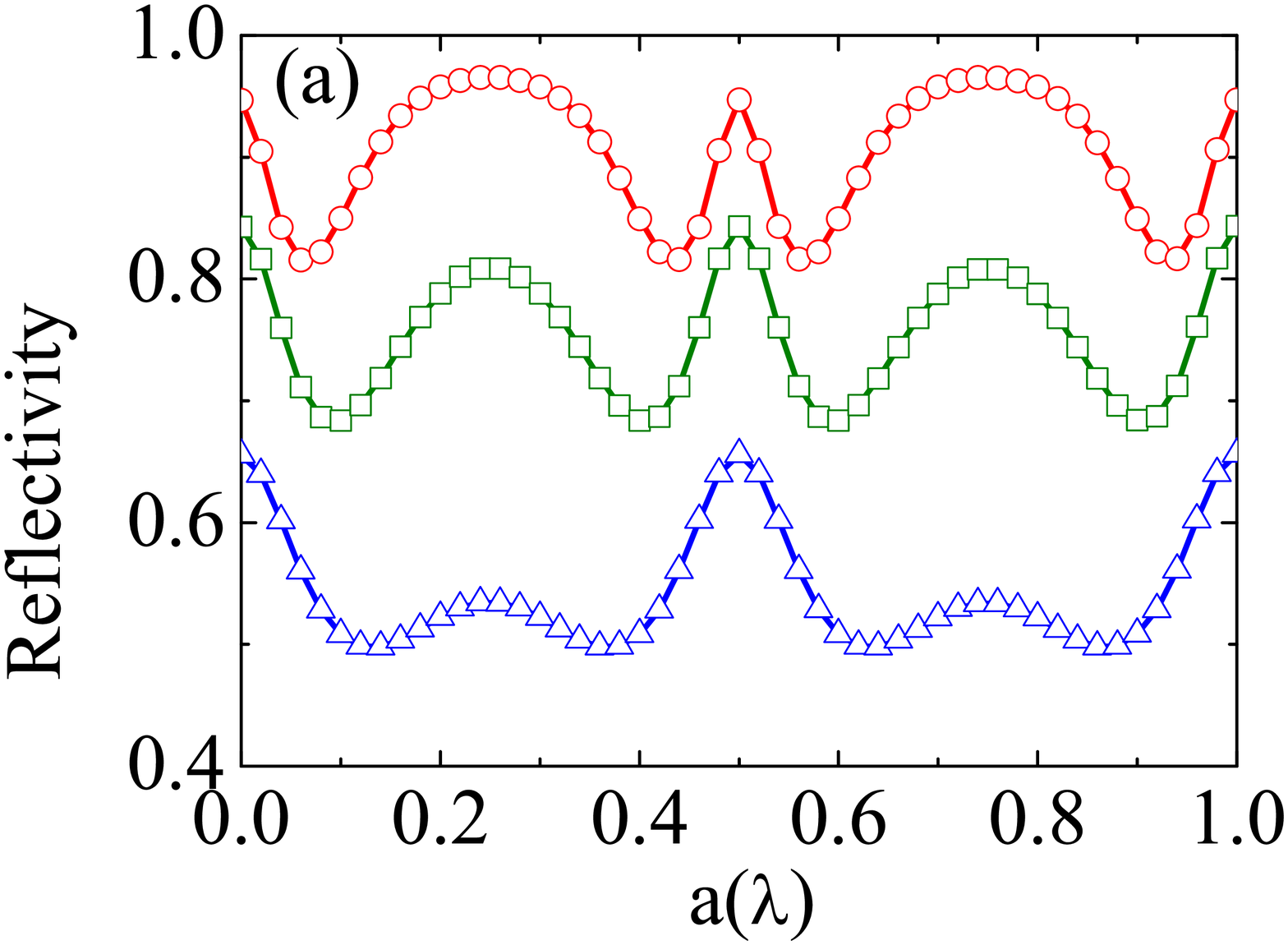}
\includegraphics[width=0.6\columnwidth]{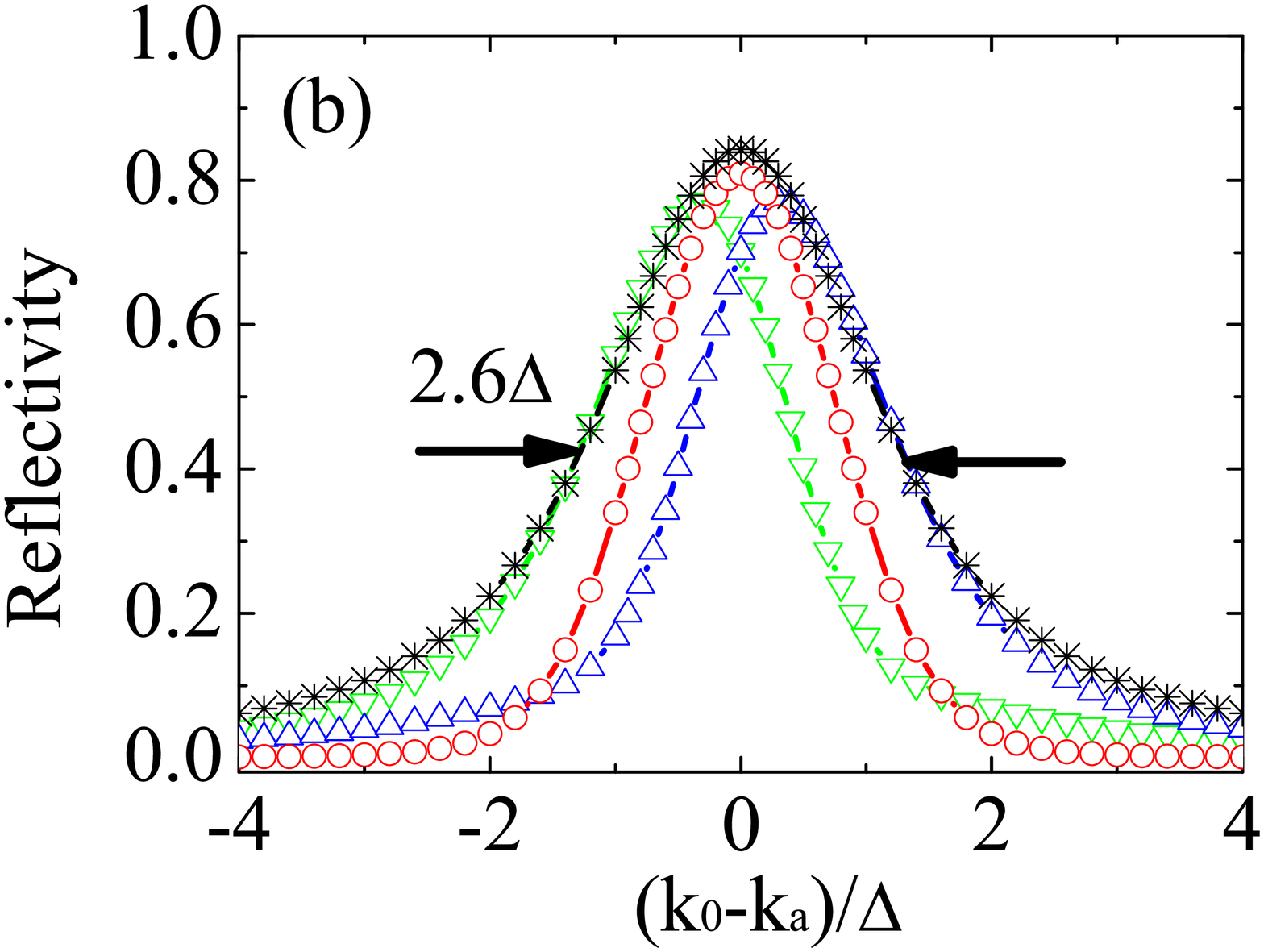}
\includegraphics[width=0.6\columnwidth]{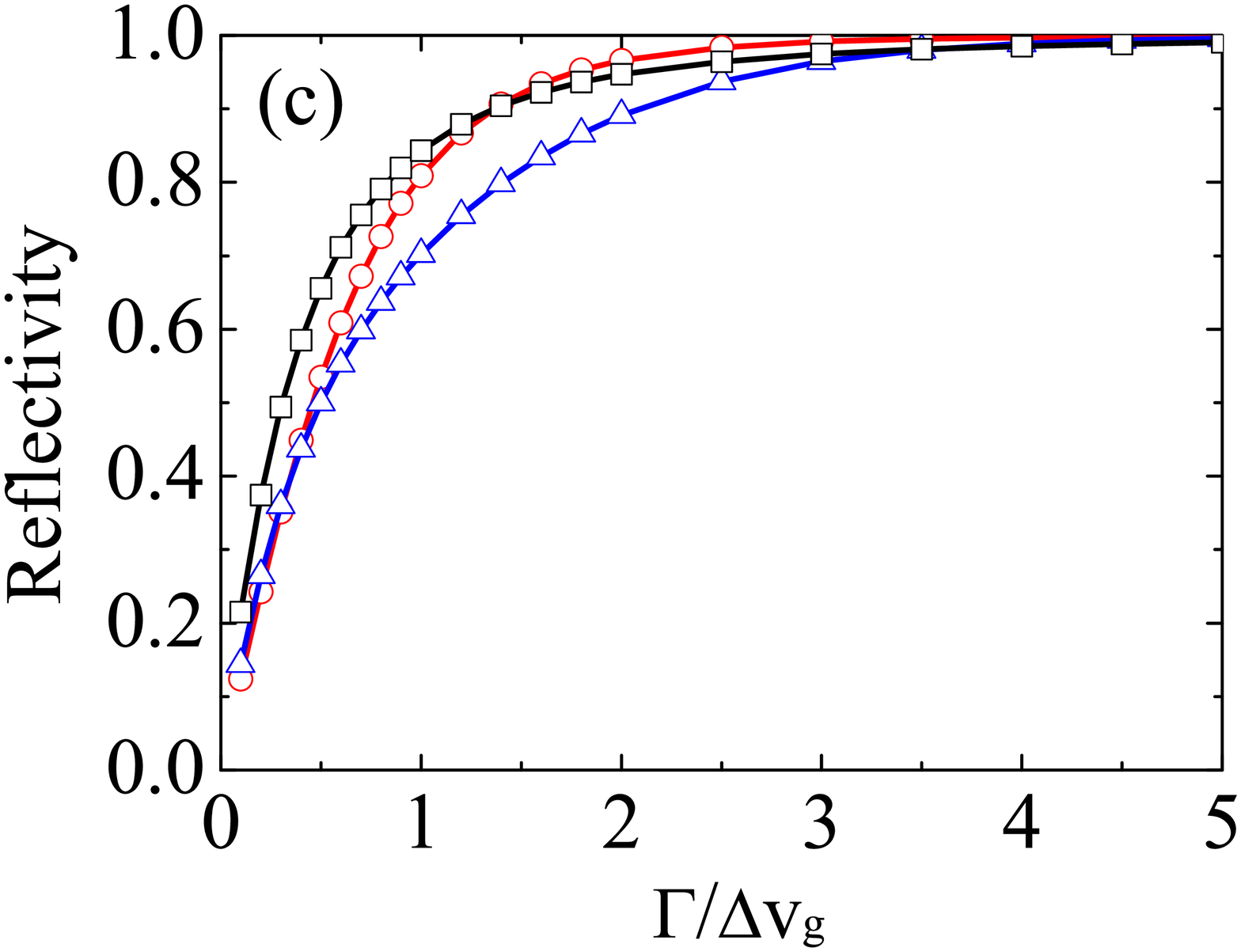}
\caption{(Color online) (a) The reflectivity as a function of atomic distance with three different coupling strengths: $\Gamma=0.5\Delta v_{g}$ (blue line with up triangles), $\Gamma=\Delta v_{g}$ (green line with squares), and $\Gamma=2\Delta v_{g}$ (red line with circles). (b) The reflectivity as a function of frequency detuning with four different atomic separations when $\Gamma=\Delta v_{g}$: $a=\lambda/2$ (black line with asterisk), $a=\lambda/4$  (red line with circles),  $a=\lambda/8$  (blue line with up triangles), and  $a=3\lambda/8$  (green line with down triangles). (c) The reflectivity as a function of $\Gamma/\Delta v_{g}$ with three different atomic separations: $a=\lambda/2$ (black line with square), $a=\lambda/4$  (red line with circles), and $a=\lambda/8$  (blue line with up triangles).}
\end{figure*}

When $a=\lambda/8$, $\lambda_{\pm}=-(\Gamma/2)[(1\pm\sqrt{2}/2)\mp i\sqrt{2}/2]$ where the collective coupling affects both the collective decay rate and the energy shift. The excitation probability as a function of time when $a=\lambda/8$ is shown in Fig. 3(g) where we see that the second atom also experiences a population oscillation due to the destructive interference. The dip in between does not touch to zero because of the unequal excitation of the two eigenmodes. The reflected and transmitted spectra is shown in Fig. 3(h) where the spectrum is very different from the previous cases. Both the reflected and transmitted spectra are asymmetric. Most of the higher frequency components are reflected while most of the lower frequency components can transmit. This can be explained by the coupling matrix. The single-photon excitation states are split into two states where $|+\rangle$ is shifted up while $|-\rangle$ is shifted down. Meanwhile, the $|+\rangle$ state has a larger coupling strength. Therefore the higher frequency components have a larger reflectivity. On the contrary, the $|-\rangle$ state has smaller coupling strength which result in a smaller reflectivity. In addition, there is a dip in the reflected spectrum due to the spontaneous emission cancellation between the two decay channel and it is a kind of Fano-like interference \cite{Fano1961, Zhu1996, Fan2002, Khanikaev2013}. The dip occurs at the position when $\delta k=\Gamma/2v_{g}$. We can also reflect the lower frequency components but transmit the higher frequencies by changing the atomic separation such that $a=3\lambda/8$. The spectra is shown in Fig. 3(i) where we can see that it is similar to the result of $a=\lambda/8$ but the spectrum is opposite. This tunability may find useful as a single photon frequency filter.

\begin{figure}
\includegraphics[width=0.8\columnwidth]{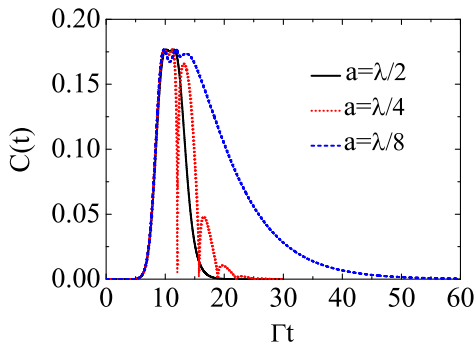}
\caption{(Color online) The concurrences as a function of time for three different atomic separations ($a=0.5\lambda$: black solid line; $a=0.25\lambda$: red dotted line; $a=0.125\lambda$: blue dashed line).  When $t=0$ the center of the input pulse is  $10/\Delta$ away from the first atom. Parameter: $\Gamma=\Delta v_{g}$.}
\end{figure}

The reflectivity as a function of atomic distance is shown in Fig. 4(a) for three different coupling strengths. We can see that the reflectivity periodically changes with the atomic separation. With small coupling strength $\Gamma=0.5\Delta v_{g}$ (blue line with up triangle), the reflectivity is maximal when the atomic distance is half integer of the wavelength. However, when we increases the coupling strength, the reflectivity increases significantly when the atomic separation is one quarter wavelength. When $\Gamma=2\Delta v_{g}$, the reflectivity with $a=\lambda/4$ is $97\%$ which is even larger than that with $a=\lambda/2$ ($95\%$). We also study how the reflectivity changes when the center frequency of the Gaussian pulse deviates from the resonant frequency. The results are shown in Fig. 4(b) where we consider four atomic separations. When $a=\lambda/4$ and $a=\lambda/2$, the reflectivity is maximum if the center frequency of the Gaussian pulse is the same as the resonant frequency. The full width at half maximum of the reflection curve when $a=\lambda/2$ is about $2.6\Delta v_{g}$ which is larger than that when $a=\lambda/4$. It is also larger than the result in single atom case which is the result of the superradiant effect. However, when $a=\lambda/8$ ($a=3\lambda/8$) the reflectivity is maximum if center frequency of the Gaussian pulse is blue (red) detuned from the resonant frequency which is due to the asymmetric coupling of the two eigenmodes. The reflectivity as a function of coupling strength for three different atomic separations are shown in Fig. 4(c). We can see that in all cases the reflectivity increases rapidly as we increase the coupling strength. The reflectivity can approach $100\%$ when $\Gamma$ is large comparing with $\Delta v_{g}$. 

Due to the collective interaction between the two atoms, quantum entanglement can also be generated in this process. After tracing over the photonic parts in Eq. (2), the bipartite quantum entanglement is calculated to be 
\begin{equation}
C(t)=\text{max}\{0,\sqrt{|\alpha_{1}(t)||\alpha_{2}(t)|}-\sqrt{2}|\alpha_{1}(t)||\alpha_{2}(t)|\}
\end{equation}
where $C(t)$ denotes the quantum concurrence of the two-atom system at time $t$ \cite{Hill1997}. The concurrences as a function of time for three different atomic separations ($a=0.5\lambda, 0.25\lambda, 0.125\lambda$) are shown in Fig. 5. For all three cases, quantum entanglement between the two atoms can be generated when the photon propagate through which is consistent with the results shown in \cite{Zheng2013}. When $a=\lambda/2$, the concurrence increases to a maximum value of about $0.17$ when the atoms are excited and stays at the maximum values for about $2/\Gamma$ before decreasing to zero (black solid curve). More interestingly, when $a=\lambda/4$ the quantum entanglement undergoes several oscillations (sudden death and revival, red dotted line) due to the interference between the two decay channels.  When $a=\lambda/8$, the quantum entanglement also undergoes small oscillations and it lasts longer than the other two cases (blue dashed line) because one of its eigenstates is sub-radiant state.

\section{N-Atom chain}

The dynamics of the atomic excitations for N-atom system with Gaussian input are given by Eq. (8) with $j=1,2,\cdots, N_{a}$. The analytical solution for many-atom system is difficult to obtain. We need to apply the numerical method to study the dynamics of the system. The spectrum of the photon after the interaction can be calculated by Eqs. (12)-(19).

The effective N-atom coupling matrix is given by
\begin{equation}
V_{N_{a}}=-\frac{\Gamma}{2}\begin{bmatrix}
1 & e^{ik_{a}a} & \cdots & e^{i(N_{a}-1)k_{a}a}\\
e^{ik_{a}a} & 1 & \cdots & e^{i(N_{a}-2)k_{a}a} \\
\vdots   & \vdots & \ddots & \vdots \\
e^{i(N_{a}-1)k_{a}a} & e^{i(N_{a}-2)k_{a}a} & \cdots & 1
\end{bmatrix}.
\end{equation}
For general cases, this matrix can not be analytically diagonalized. However, when $a=\lambda/2$, the matrix can be diagonalized and the eigenvalues are given by $\lambda_{1}=N_{a}\Gamma/2$ and $\lambda_{2,\ldots, N_a}=0$. Only one eigenmode can couple to the waveguide modes and the coupling strength is $N_{a}$ times the single atom coupling strength which is the collective superradiant effect. The superradiant eigenstate is given by $(1/\sqrt{N_{a}})\sum_{i=1}^{N_{a}}(-1)^{i}|g,\ldots,e_{i},\ldots,g\rangle$ which is the timed Dicked state \cite{Scully2006}.
For general cases, multiple eigenstates can couple to the ground state and they may interfere with each other. 

\begin{figure*}
\includegraphics[width=0.6\columnwidth]{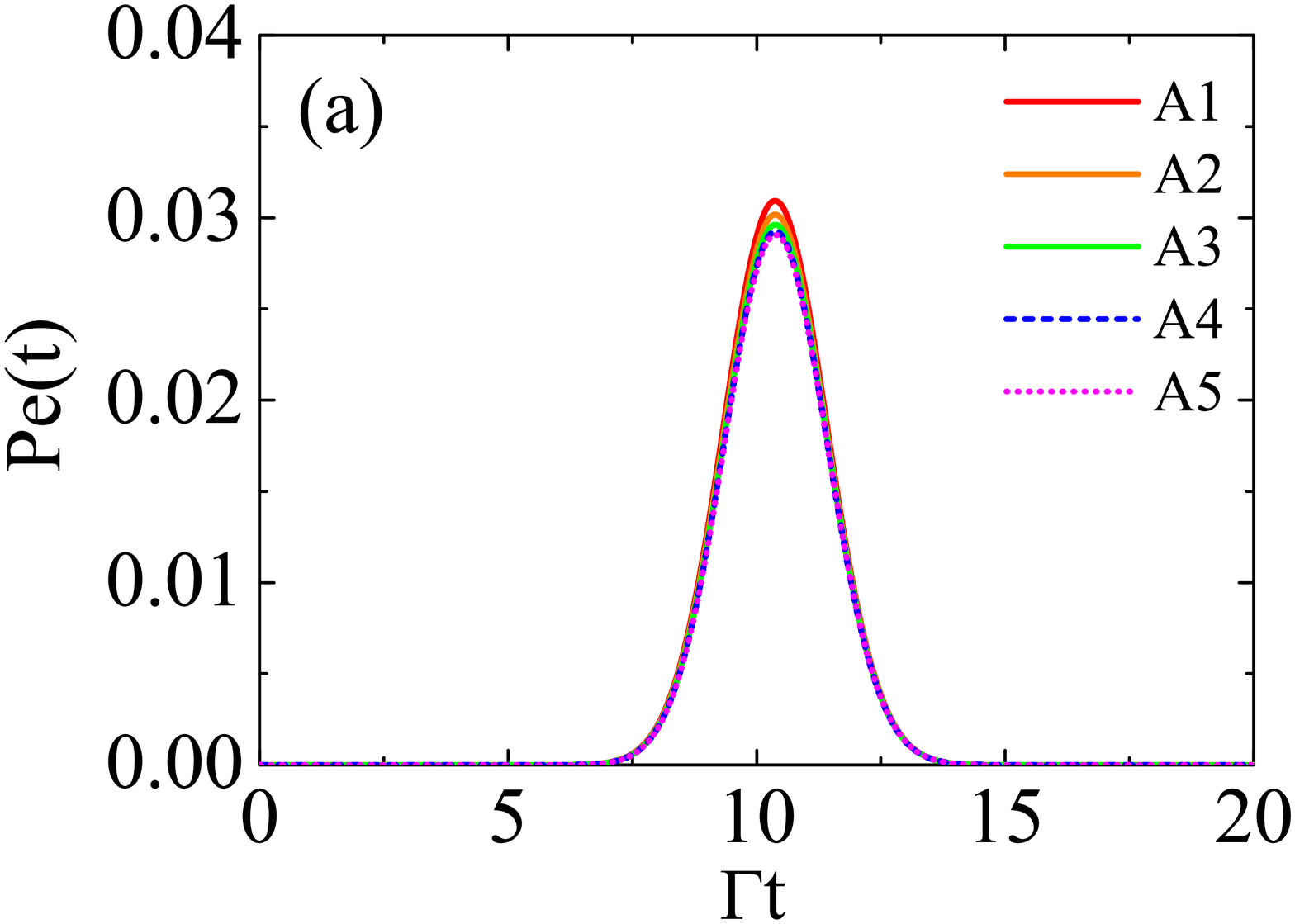}
\includegraphics[width=0.6\columnwidth]{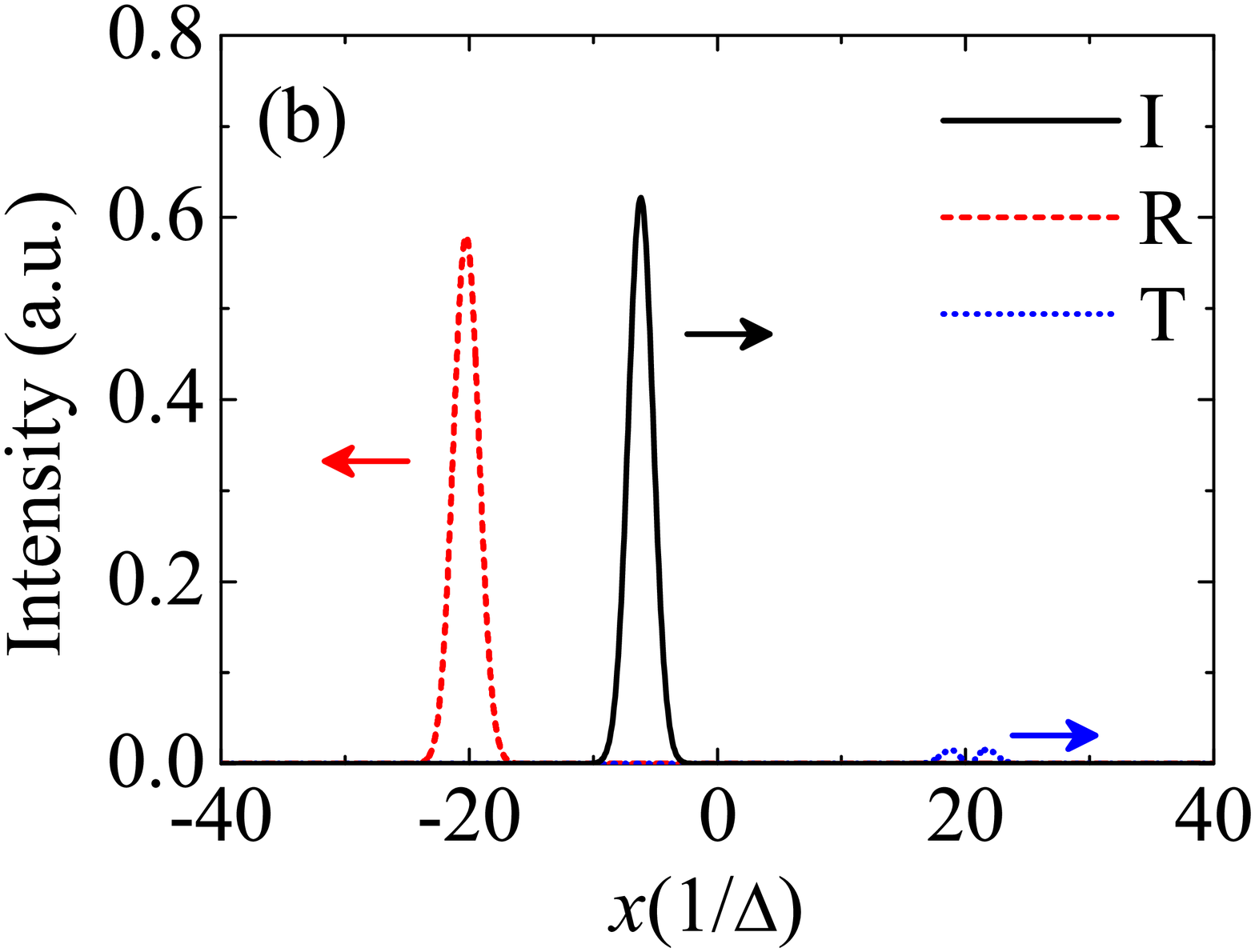}
\includegraphics[width=0.6\columnwidth]{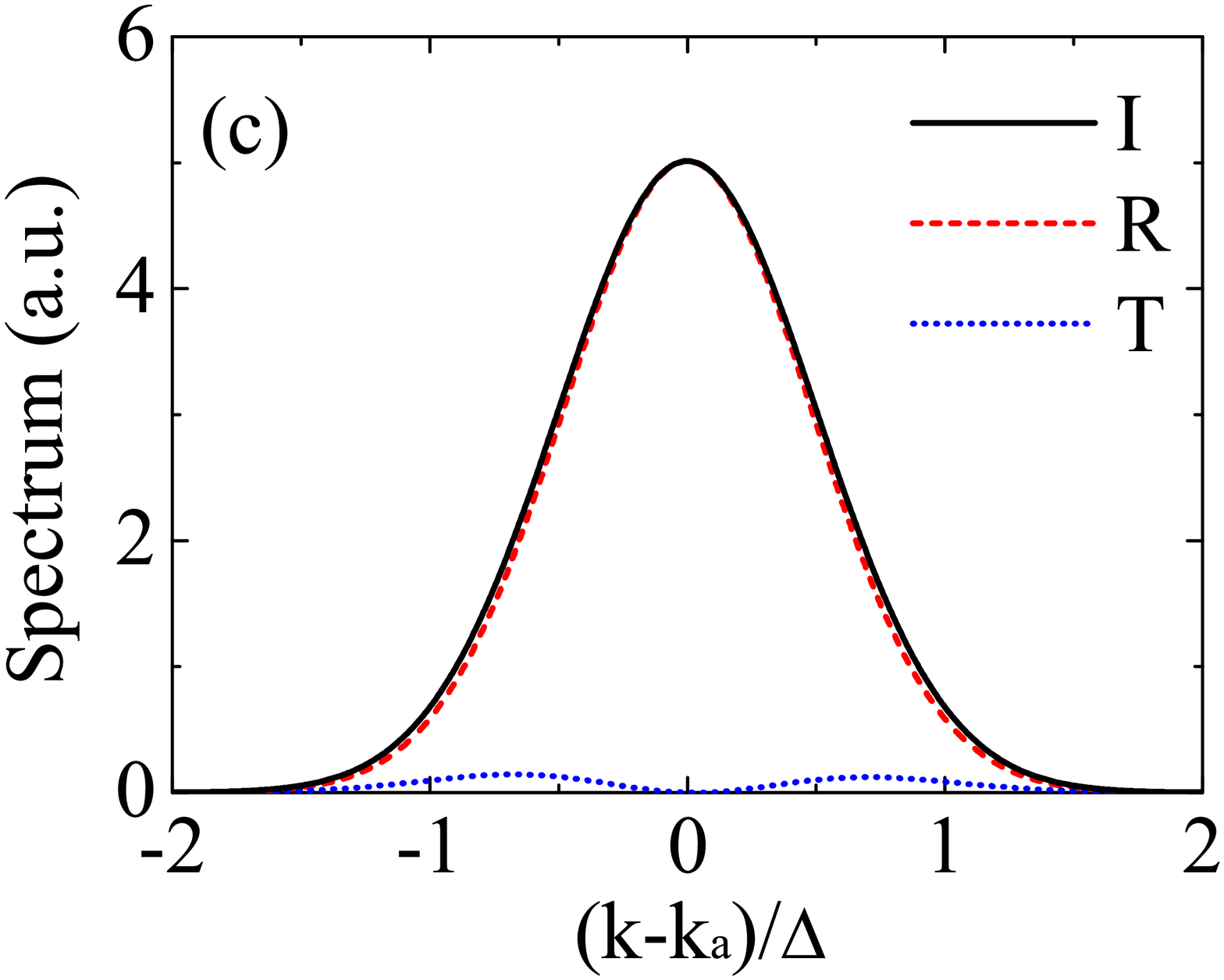}
\includegraphics[width=0.6\columnwidth]{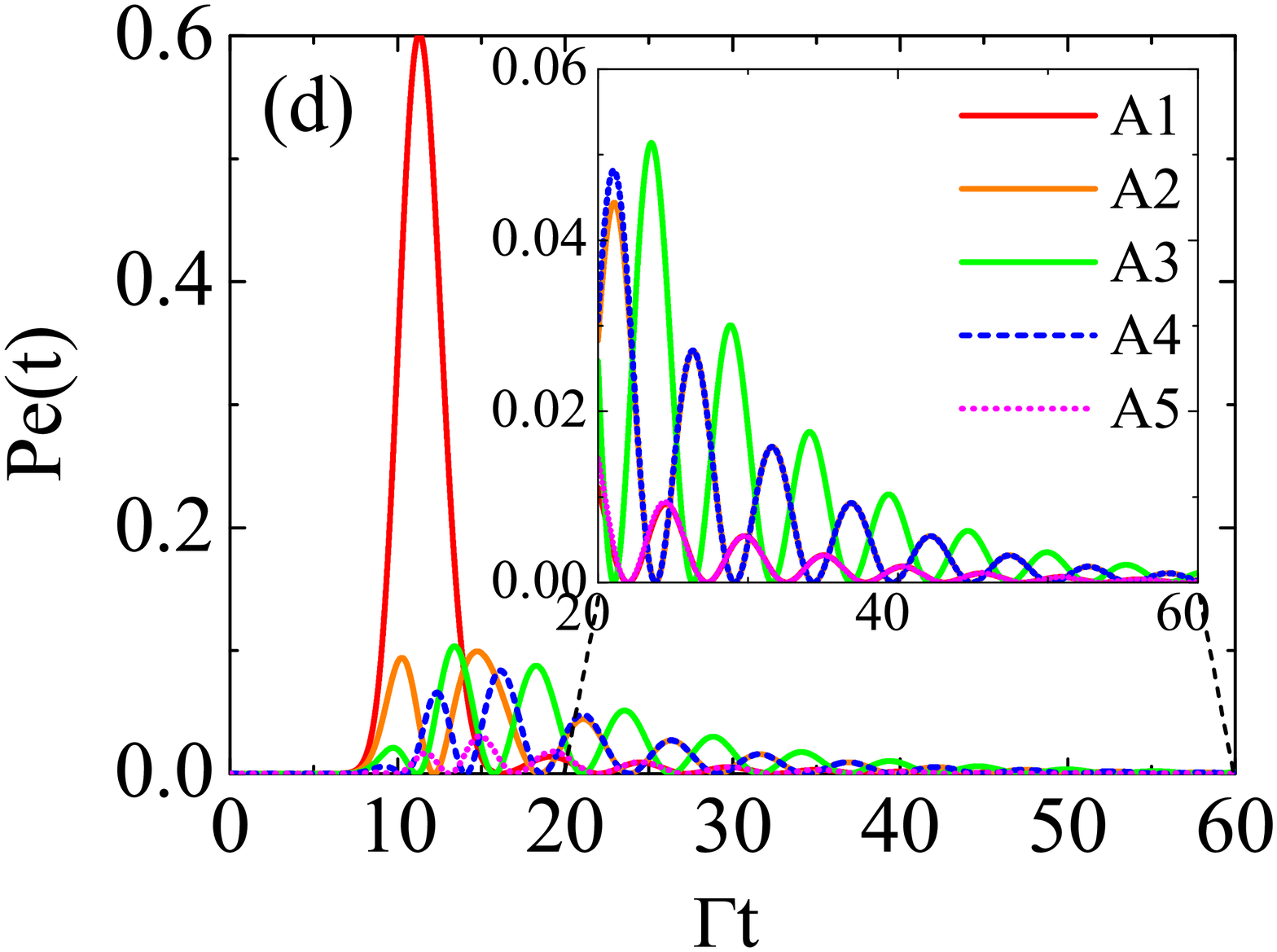}
\includegraphics[width=0.6\columnwidth]{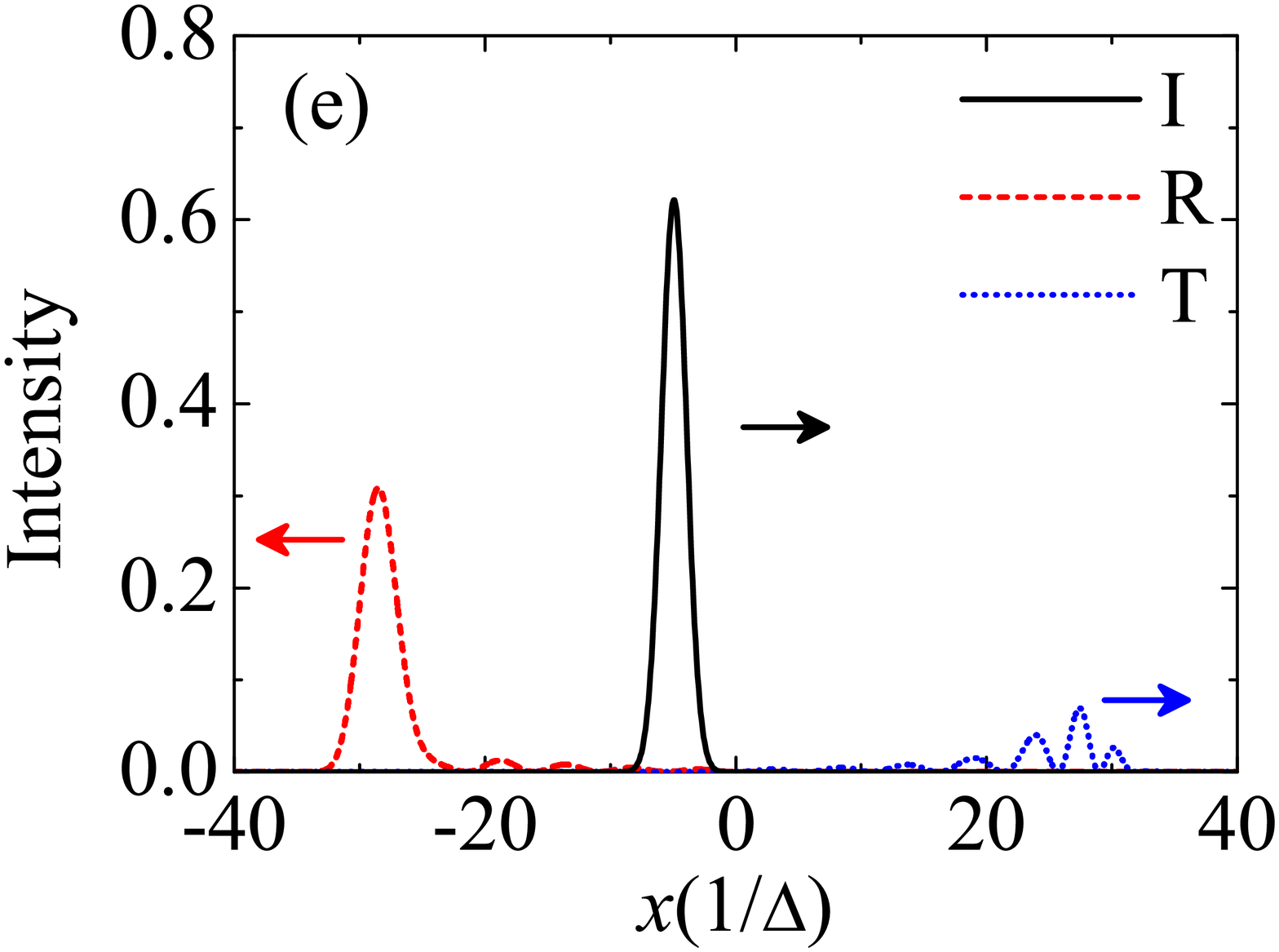}
\includegraphics[width=0.6\columnwidth]{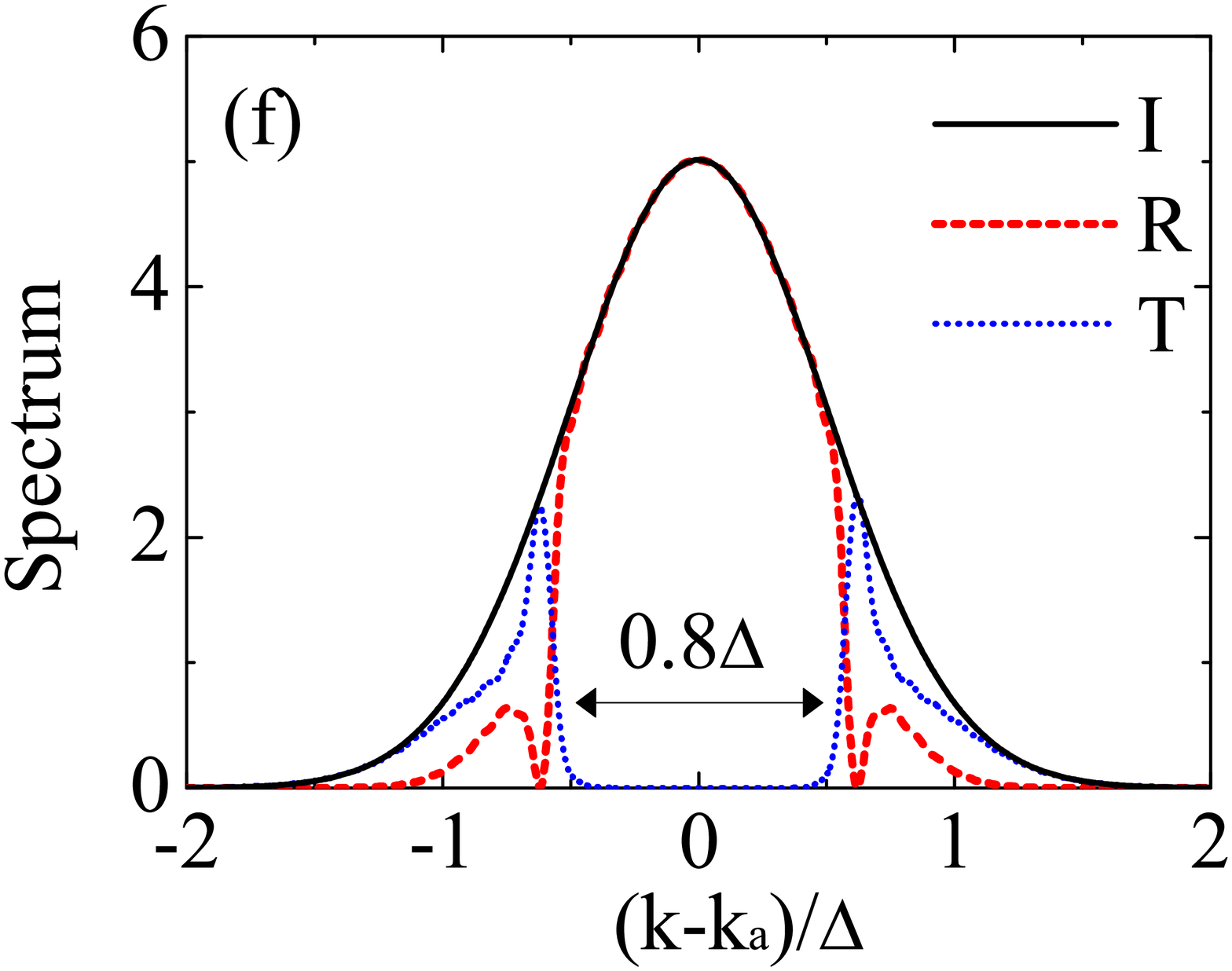}
\includegraphics[width=0.6\columnwidth]{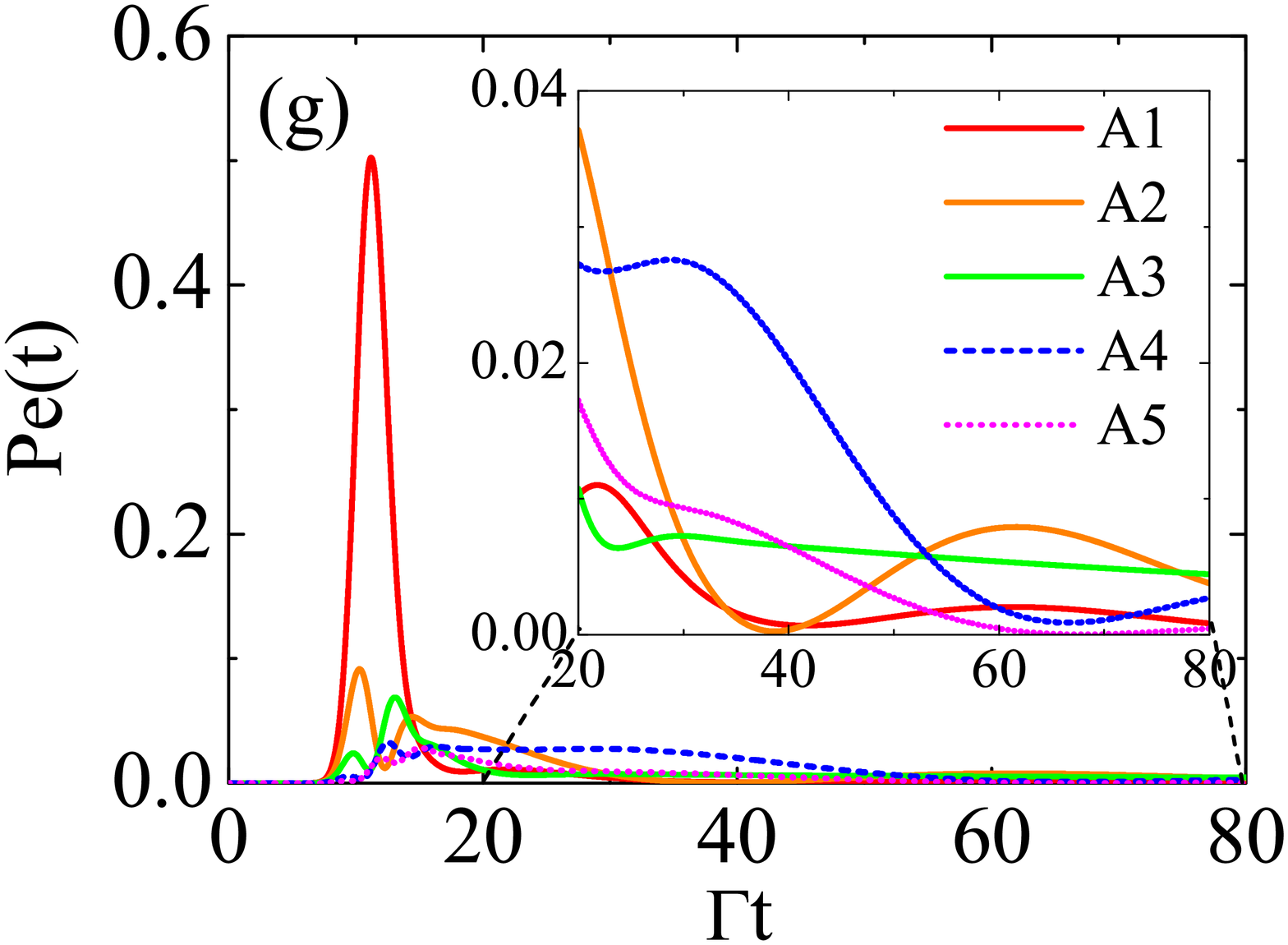}
\includegraphics[width=0.6\columnwidth]{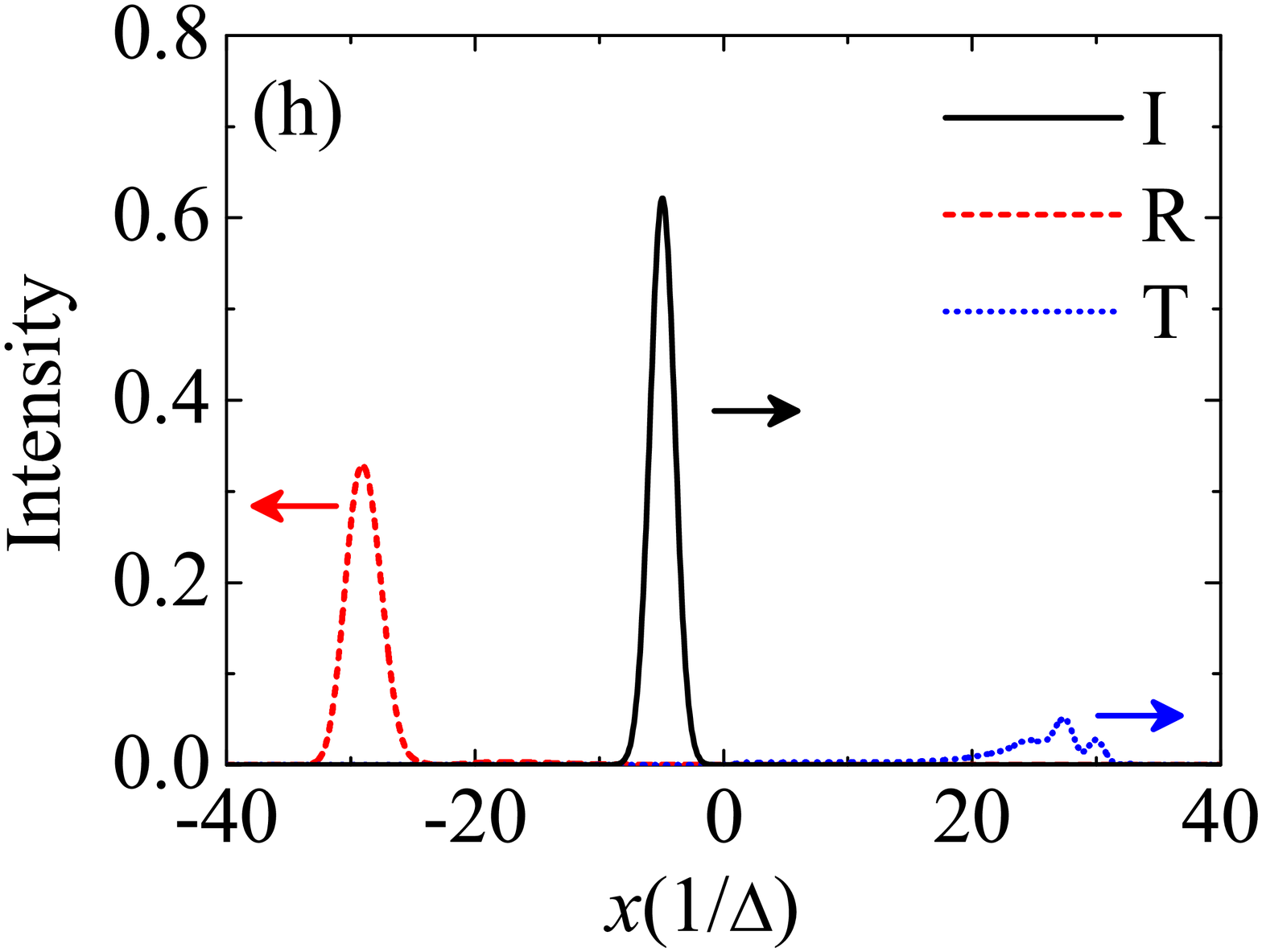}
\includegraphics[width=0.6\columnwidth]{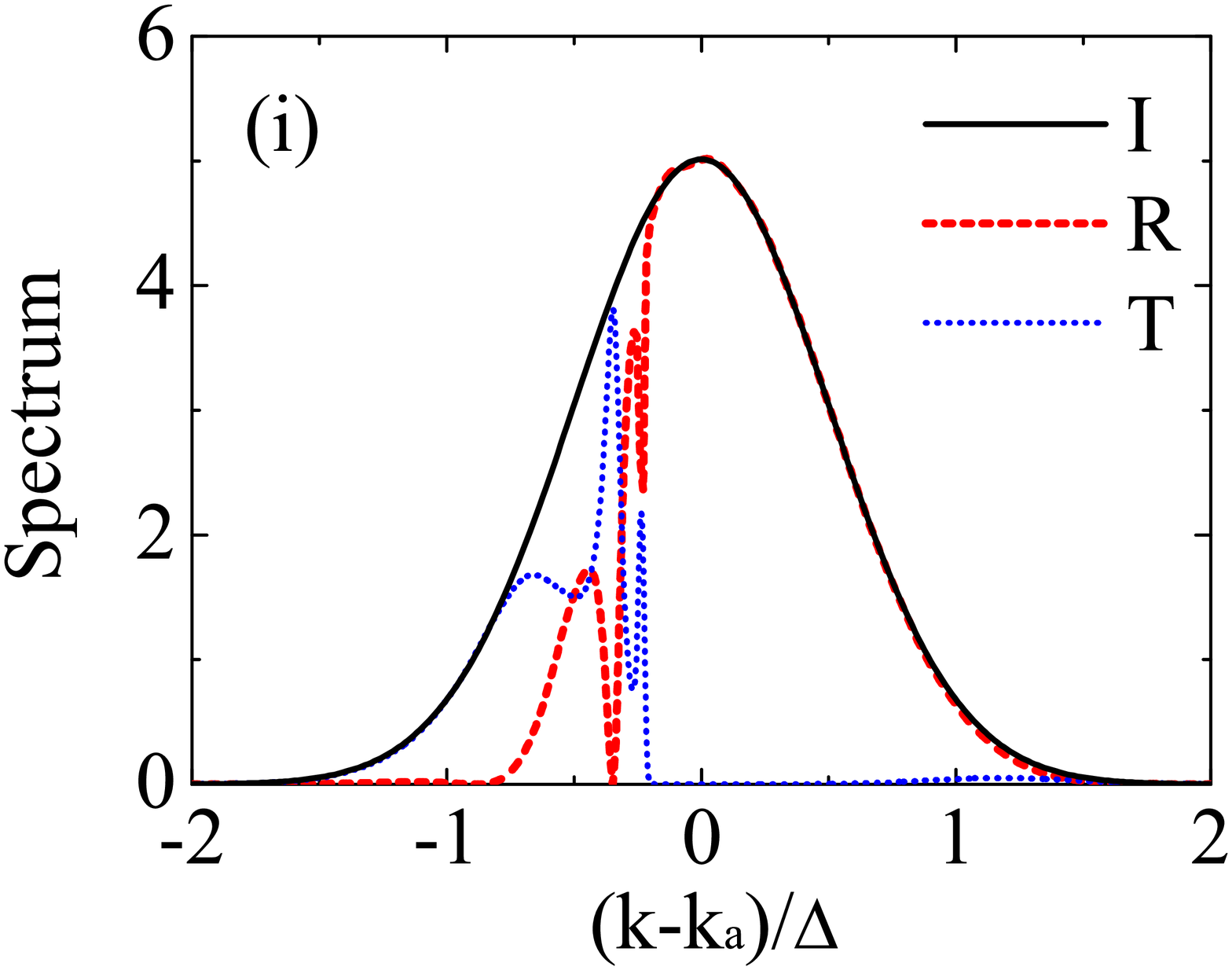}
\caption{(Color online) (a,d,g) The atomic excitation probabilities as a function of time when the photon passes through. When $t=0$ the center of the input pulse is  $10/\Delta$ away from the first atom. Here Ai is for the ith atom. (b,e,h) The pulse shapes of the incoming (I, at $t=6/\Gamma$), reflected (R, at $t=30/\Gamma$) and transmitted (T, at $t=30/\Gamma$) photon. (c,f,i) The spectrum (arb. units) of the incoming (I), reflected (R) and transmitted (T) photon. Parameters: $\Gamma =\Delta v_g$. (a-c) $a=\lambda/2$, (d-f) $a=\lambda/4$, (g-i) $a=\lambda/8$.}
\end{figure*}

\begin{figure*}
\includegraphics[width=0.7\columnwidth]{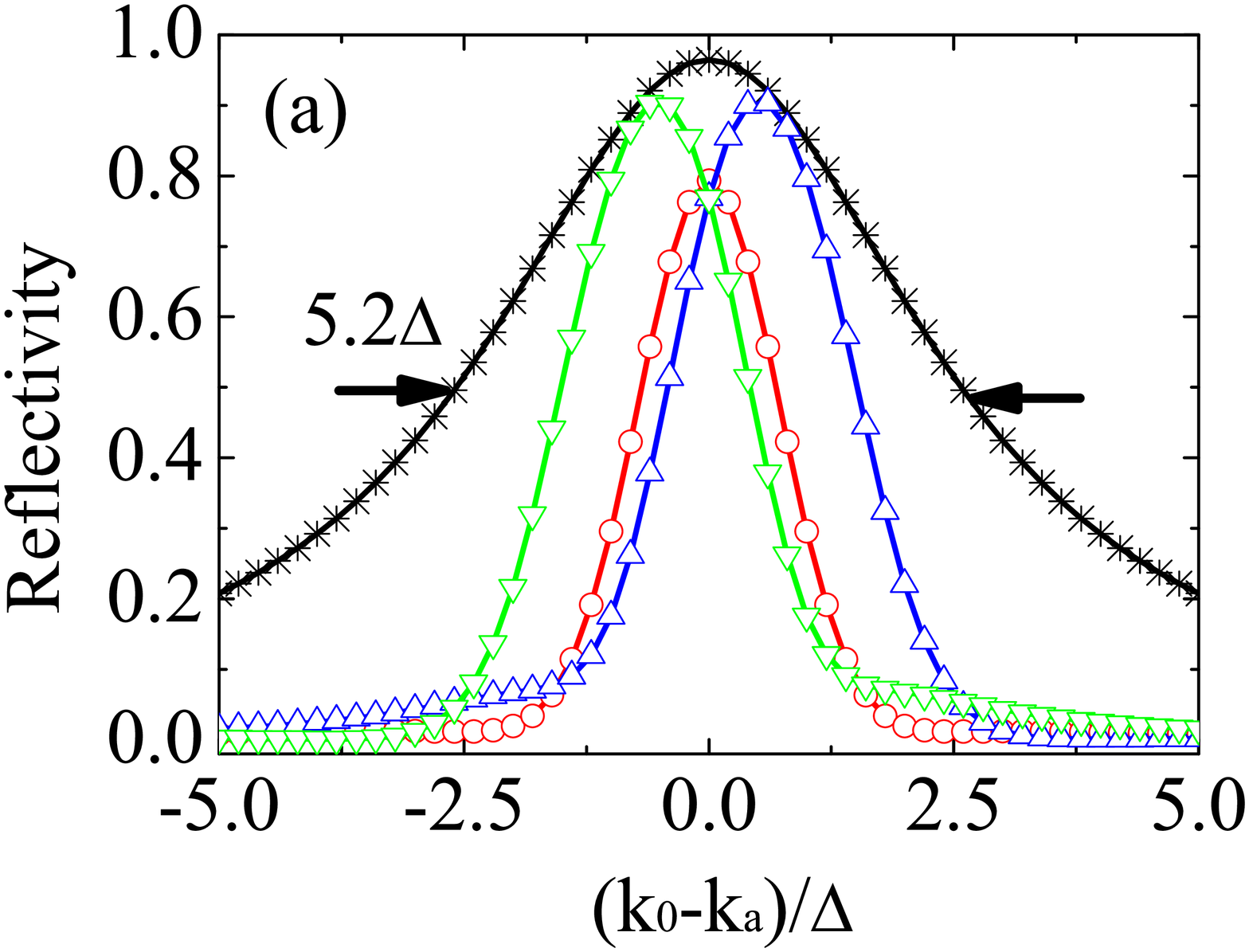}
\includegraphics[width=0.7\columnwidth]{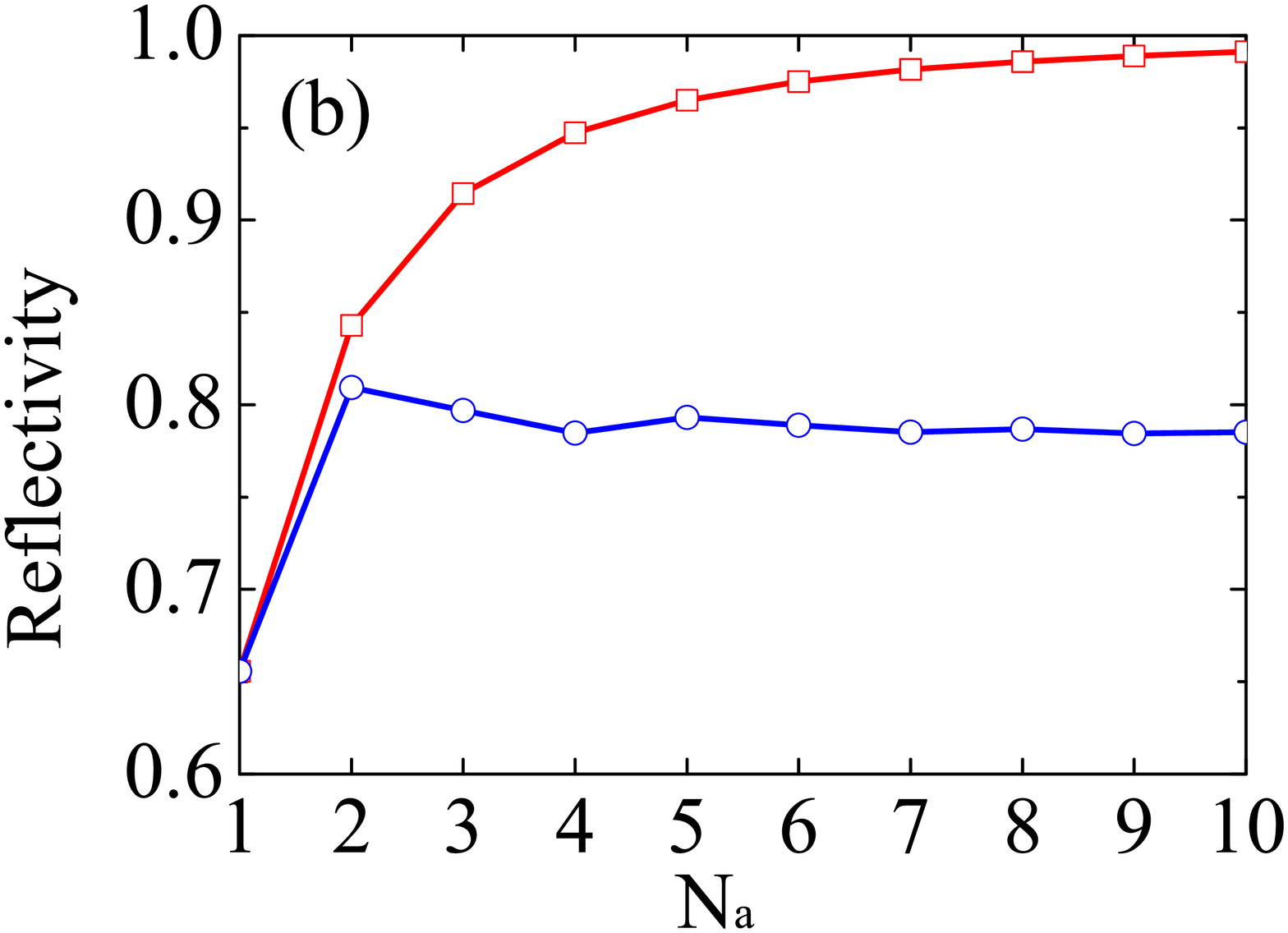}
\caption{(Color online) (a) The reflectivity as a function of center frequency deviation for different atomic separations: $a=\lambda/2$ (black line with asterisk), $a=\lambda/4$  (red line with circles),  $a=\lambda/8$  (blue line with up triangles), and  $a=3\lambda/8$  (green line with down triangles). (b) The reflectivity as a function of atom numbers for two atomic separations: $a=\lambda/2$ (red line with squares) and $a=\lambda/4$ (blue line with circles). Parameters: $\Gamma=\Delta v_{g}$.}
\end{figure*}

Let us take $N_{a}=5$ as an example to study the behaviors of the photon transport in this multiple-atom system. The results when $a=\lambda/2$ and $\Gamma=\Delta v_{g}$ are shown in Fig. 5(a-c) where (a) is the excitation probability for each atom as a function time when the photon passes through, (b) is the photon pulse before and after the interaction and (c) is the corresponding spectra. In this case, only the superradiant eigenstate is coupled with the guided photon and we can prepare the system in a timed Dicke state where each atom has an equal probability to be excited. From Fig. 5(a), we can see that all of the atoms are indeed excited and deexcited at almost the same rate. The small differences in each amplitude come from the time-retarded effect. The reflected and transmitted photon pulse shape and spectrum are similar to the single atom case because the collective coupling only affects the collective decay but not the energy shift when $a=\lambda/2$. Since the collective coupling and decay rate here is 5 times larger than that of the single atom case, the reflectivity here is much larger which is about $96\%$ comparing with $66\%$ in the single atom case.

The results when $a=\lambda/4$ and $\Gamma=\Delta v_{g}$ are shown in Fig. 5(d-f). The atomic excitation probability as a function of time is shown in Fig. 5(d) where we can see that the five atoms have quite different excitation dynamics and they oscillate with time due to the interferences between each coupling channel. In addition, the atomic excitations last much longer than one atom case because some photon energy are trapped in the subradiant states of the system. The reflected and transmitted pulses are also quite different here (Fig. 5(e)). Both the reflected and transmitted pulses have multiple peaks and valleys due to the interferences of the reflected and transmitted photon amplitude. A finite bandwidth of frequencies around the resonant frequency are almost completely reflected which is the demonstration of photonic bandgap effect (Fig. 5(e)). The reflection and transmission window here is about $0.8\Delta v_{g}$ which is larger than that in two-atom case and it can also be tuned by changing the coupling strength which may be used as a single photon frequency filter. Different from the two-atom cases, the reflection spectrum here has two dips due to the Fano-like interferences between the multiple decay channels \cite{Fano1961, Fan2002, Khanikaev2013}.

The results when $a=\lambda/8$ and $\Gamma=\Delta v_{g}$ are shown in Fig. 5(g-i). Similar to the case when $a=\lambda/4$, the atomic excitation can exist for a very long time and the excitation probability can oscillate with time (Fig. 5(g)). However, the oscillation period here is longer than that when $a=\lambda/4$. The reflected and transmitted photon shapes are shown in Fig. 5(h). The spectra are shown in Fig. 5(i) where we see that most of the higher frequency components are reflected and most of the lower frequency components can transmit. This property may be used as a single photon frequency filter. There are multiple dips in the reflected spectrum due to Fano-like interferences between multiple eigenstates \cite{Fano1961, Fan2002, Khanikaev2013}. 

\begin{figure*}
\includegraphics[width=0.6\columnwidth]{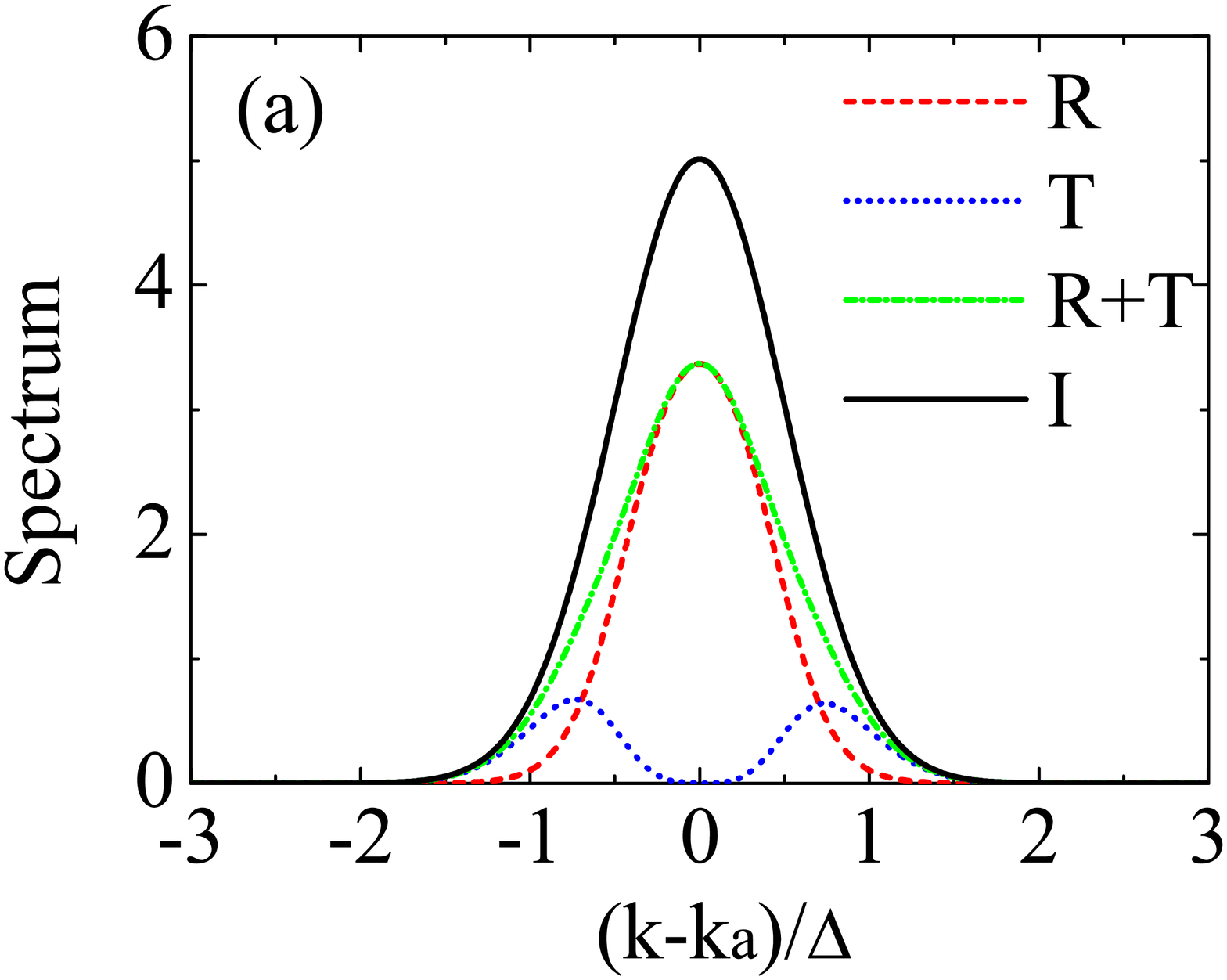}
\includegraphics[width=0.6\columnwidth]{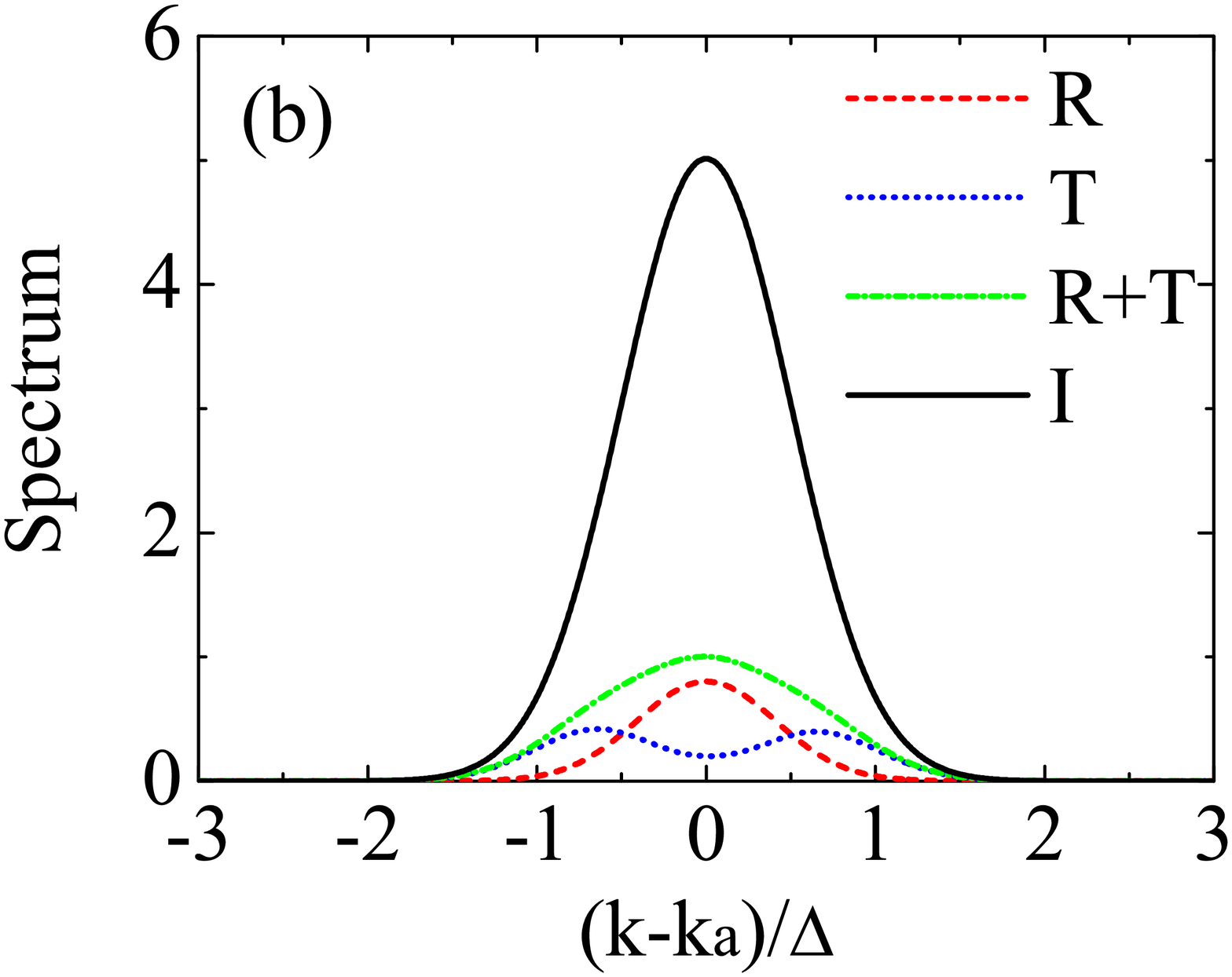}
\includegraphics[width=0.6\columnwidth]{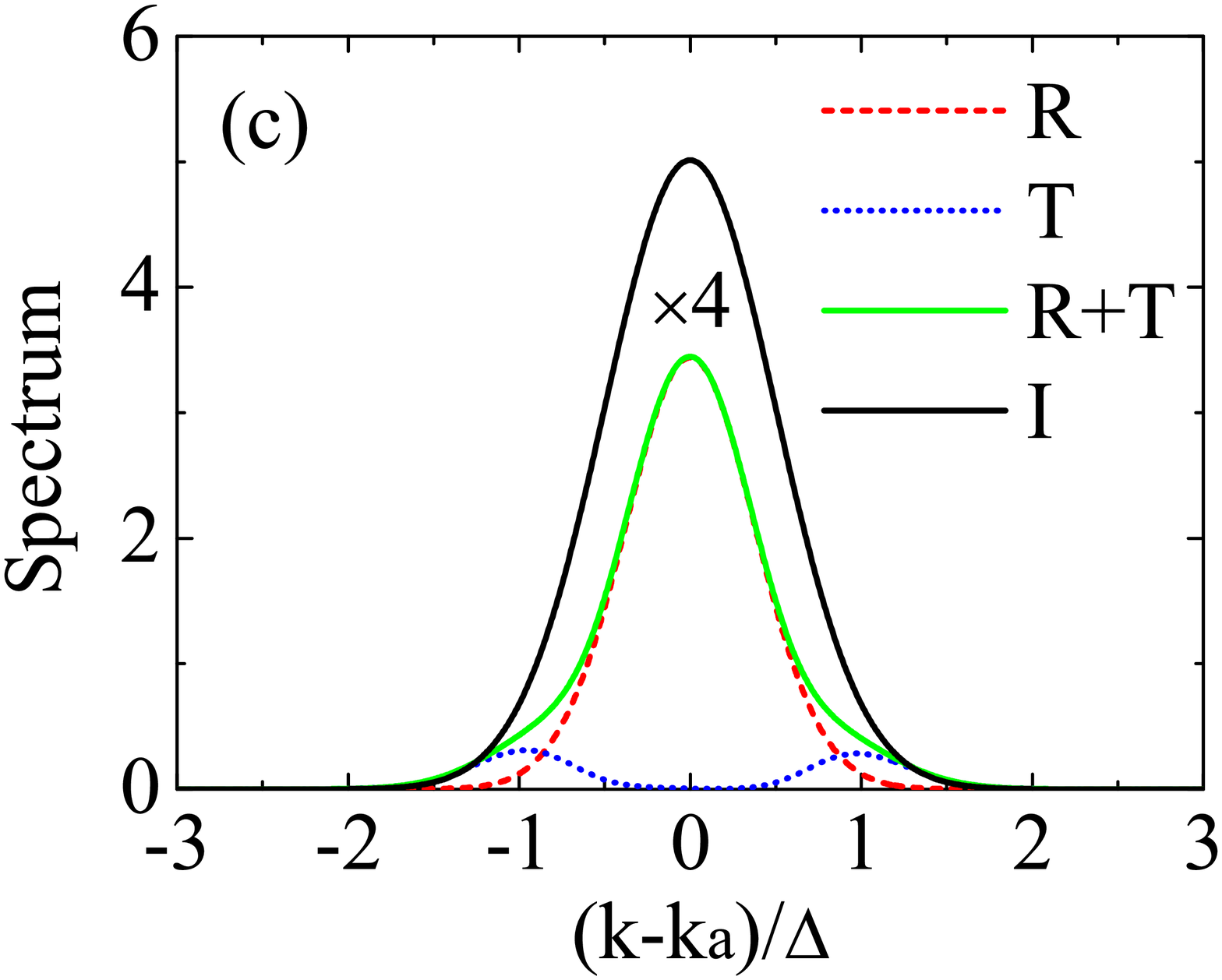}
\caption{(Color online) The spectrum (arb. units) of the input (I), reflected (R) and transmitted (T) photon with spontaneous decay to the free space. The black solid line is the input spectrum. The red dashed line is the spectrum of the reflected photon. The blue dotted line is the spectrum of the transmitted photon. The green dash-dotted line is the total spectrum of the photon remains in the waveguide.  (a) $N=2$, $\Gamma=\Delta v_{g}$, and $\gamma=\Gamma/5$; (b) $N=2$, $\Gamma=\Delta v_{g}$, and $\gamma=\Gamma$; (c) $N=5$, $\Gamma=\Delta v_{g}$, and $\gamma=\Gamma$. }
\end{figure*}

The reflectivity as a function of center frequency deviation for four different atomic separations are shown in Fig. 6(a). Similar to the two-atom cases, the reflectivity is maximum if the center frequency of input photon is resonant with the atomic transition frequency for the cases when $a=\lambda/4$ and $a=\lambda/2$, while the reflectivity is maximum if the center frequency of input photon is blue or red detuned from the resonant frequency for the cases when $a=\lambda/8$ and $a=3\lambda/8$ due to asymmetric coupling. The full width at half maximum of the reflectivity curve when $a=\lambda/2$ is about $5.2\Delta v_{g}$ which is much broader than the one-atom and two-atom cases which is a superradiant effect of this collective many-body system \cite{Dicke}. 

Finally, we study how reflectivity changes as we increase the atom numbers in the chain. The results are shown in Fig. 6(b) where we consider two atomic separations, i.e., $a=\lambda/2$ and $a=\lambda/4$. When $a=\lambda/2$, the reflectivity increases as the atom number increases and the reflectivity can approach $100\%$ if the atom number is large enough. However, when $a=\lambda/4$ the reflectivity does not significantly change when we increase the atom number. Therefore, to effectively reflect a photon, one way is by increasing the coupling strength and the other way is by increasing the collective coupling. To increase the collective coupling, we can increase the atom numbers but the separation between atoms need to be half integer of the wavelength. 

\section{Effects of the spontaneous decay}

In previous sections, we have neglected the effects of the spontaneous decay to the free space. In this section we numerically study how the spontaneous decay of the atom to the free space affects the scattering properties. Let us take $N=2$ atom for example. For two-atom system, Eq. (14) including $\gamma$ becomes
\begin{equation}
-i\delta kv_{g}\vec{\chi}(\delta k)=\vec{b}(\delta k)-\frac{\gamma}{2} \vec{\chi}(\delta k) +V_{2}(\delta k)\vec{\chi}(\delta k)
\end{equation}
with $\vec{\chi}=(\chi_{1},\chi_{2})^T$, $\vec{b}=(b_{1},b_{2})^T$. It is readily to solve for $\chi_{1}$ and $\chi_{2}$ from Eq. (43). The spectra after the interaction can then be calculate from Eqs. (18) and (19) and they are given by
\begin{eqnarray}
\beta^{R}_{\delta k}&=&\beta_{\delta k}(0)\frac{\gamma^{2}-4i\delta kv_{g}\gamma-4\delta k^{2}v_{g}^{2}}{(\Gamma+\gamma-2i\delta kv_{g})^2-\Gamma^{2}e^{2ika}}, \\
\beta^{L}_{\delta k}&=&\beta_{\delta k}(0)e^{2ikr_{1}}\frac{(1+e^{2ika})(\Gamma+\gamma-2i\delta k v_{g})-2\Gamma e^{2ika}}{(\Gamma+\gamma-2i\delta kv_{g})^2-\Gamma^{2}e^{2ika}}, \nonumber \\
\end{eqnarray} 
where $k=k_{a}+\delta k$.

The results are shown in Fig. 8(a) and (b) where the black solid line is the spectrum of the input photon, the red dashed line is the spectrum of the reflected photon, the blue dotted line is the spectrum of the transmitted photon, and the green solid line is the summation of the reflected and transmitted photon in the waveguide. When there are only two atoms, the result for $\gamma=\Gamma/5$ is shown in Fig. 8(a) and the result if $\gamma=\Gamma$ is shown Fig. 8(b). In both cases, the total probability that the photon stays in the waveguide is reduced because the photon can leak out to the free space. However, we notice that the general features of the scattering spectrum do not change for smaller decay rate when $\gamma=\Gamma/5$, i.e., there is a range of spectrum that can not transmit. In this case the photon has a probability about $68\%$ staying in the guided modes. For larger decay rate when $\gamma=\Gamma$, only about $26\%$ probability the photon can stay in the waveguide and the general features change where the photonic bandgap effect disappears. This is because the spontaneous decay can eliminate the interference of the emitted photon from the two atoms. However, if we increase the atom number such that $N_{a}=5$ the photonic bandgap effect appears again even if $\gamma=\Gamma$ (Fig. 8(c)). This is because the collective coupling can be much larger than the spontaneous decay rate.

Here, we should note that we have neglected the dipole-dipole interaction between the atoms induced by the vacuum modes in the free space. This is a good approximation if the atomic separation is not very small (e. g. larger than $\lambda/10$) \cite{Ficek2004, Liao2012}. However, if the atomic separation is very small (e. g., smaller than $\lambda/10$), the dipole-dipole interaction between the atoms induced by the vacuum modes can be comparable with or even larger than the spontaneous decay rate $\gamma$ and it should be included in the calculations. In this paper, we focus on the dipole-dipole interaction between the atoms induced by the guided modes which is a long-range effect and the smallest atomic separation used in the numerical calculations is $\lambda/8$ where the dipole-dipole interaction induced by the vacuum modes in the free space can be safely neglected.

\section{Summary}  

In summary, we derive a time-dependent theory to study the dynamics of a single photon pulse propagates through a linear atomic chain which is coupled to a 1D single mode photonic waveguide. This theory allows us to study the real time evolution of a single photon transport and atomic excitations. The shape of the single photon pulses and the arrangements of the atoms can be both arbitrary. Our result is consistent with the previous calculations when there is only one atom. For more atoms, the collective interactions can significantly modify the dynamics of the system. The reflectivity can approach $100\%$ for a photon with finite bandwidth by changing the ratio of coupling strength and the photon linewidth or by changing the number of atoms in the chain to increase the collective coupling strength. The spectrum of the reflected and transmitted photon can also be significantly different from that of the single atom case. We numerically show the photonic band gap effects where a bandwidth of frequency can not transmit and we can also tune this gap by changing $\Gamma/\Delta v_{g}$ or by changing the number of atoms. We can also reflect certain part of the frequency components but transmit the other part of frequency by controlling the collective coupling between the atoms. This may find important applications for single photon frequency filter. The quantum entanglement between distant atoms can also be created by the guided photon which may find important applications in quantum information and quantum computing. Besides, we also demonstrate the Fano-type interference and the supperradiant effects in this system.

The spontaneous decay to the free space can reduce the photon amplitude that stays in the photonic waveguide. If the coupling strength between the atom and the waveguide mode is much larger than the spontaneous decay rate, the general feature is similar to the case when we neglect the spontaneous decay. However, when the spontaneous decay rate is comparable to  the coupling strength between the atom and the waveguide mode, the feature of the reflected and transmitted spectrum can be quite different. However, if we increase the atom number to increase the collective coupling strength, the reflected and transmitted spectrum can have feature similar to the case without spontaneous decay rate.

\section{Acknowledgment}
This work is supported by a grant from the Qatar National Research Fund (QNRF) under the NPRP project 7-210-1-032. S. -Y. Z. is supported by the Joint Fund of the National Natural Science Foundation of China and the China Academy of Engineering Physics (No. U1330203).

\section*{APPENDIX: CONCURRENCE OF TWO-QUBIT MIXED STATE}

The quantum entanglement of a two-qubit system can be well quantified by quantum concurrence which is given by \cite{Hill1997}
\begin{equation}
C(\rho )=\text{max}\{0,\lambda_1-\lambda_2-\lambda_3-\lambda_4 \}
\end{equation}
where $\lambda_i (i=1,2,3,4)$ are the eigenvalues, in decreasing order, of the Hermitian Matrix $\Lambda =\sqrt{\sqrt{\rho }\tilde{\rho}\sqrt{\rho} }$ with $\tilde{\rho}=(\sigma_{y}\otimes \sigma_{y})\rho^{*}(\sigma_{y}\otimes \sigma_{y})$.

For the two-atom system in this paper, we have
\begin{equation}
\rho_{AF}(t)=|\Psi(t)\rangle \langle \Psi(t)|
\end{equation}
with 
\begin{equation}
|\Psi(t)\rangle =\sum_{j=1}^{2}\alpha_{j}(t)e^{-i\omega_{a}t}|e_{j},0\rangle+\sum_{k}\beta_{k}(t)e^{-i\omega_{k}t}|g,1_{k}\rangle. 
\end{equation}
The density matrix of the atomic system is $\rho_{A}(t)=Tr_{F}(\rho_{AF}(t))$ which can be calculated to be
\begin{equation}
\rho_{A}(t)=\begin{bmatrix}
\sum_{k}|\beta_{k}(t)|^2 & 0 & 0 & 0\\
0 & |\alpha_{1}(t)|^{2} & \alpha_{1}(t)\alpha_{2}^{*}(t) & 0 \\
0   & \alpha_{1}^{*}(t)\alpha_{2}(t) & |\alpha_{2}(t)|^2 & 0 \\
0 & 0 & 0 & 0
\end{bmatrix}.
\end{equation}

The Hermitian matrix $\Lambda(t)$ is then given by
\begin{widetext}
\begin{equation}
\Lambda(t)=\begin{bmatrix}
0 & 0 & 0 & 0\\
0 & |\alpha_{1}(t)|^{2}|\alpha_{2}(t)|^{2} & |\alpha_{1}(t)||\alpha_{2}(t)|\alpha_{1}(t)\alpha_{2}^{*}(t) & 0 \\
0   & |\alpha_{1}(t)||\alpha_{2}(t)|\alpha_{1}^{*}(t)\alpha_{2}(t) & |\alpha_{1}(t)|^{2}|\alpha_{2}(t)|^2 & 0 \\
0 & 0 & 0 & 0
\end{bmatrix}
\end{equation}
\end{widetext}
and its eigenvalues are given by $\lambda_{1}(t)=\sqrt{|\alpha_{1}(t)||\alpha_{2}(t)|}, \lambda_{2}(t)=\sqrt{2}|\alpha_{1}(t)||\alpha_{2}(t)|,\lambda_{3}(t)=\lambda_{4}(t)=0$. Hence, the concurrence of the two-atom system is given by
\begin{equation}
C(t)=\text{max}\{0,\sqrt{|\alpha_{1}(t)||\alpha_{2}(t)|}-\sqrt{2}|\alpha_{1}(t)||\alpha_{2}(t)|\}
\end{equation}
which is the Eq. (41).

\end{document}